\documentclass[nofootinbib,floatfix,superscriptaddress,twocolumn]{revtex4}
\pdfoutput=1
\usepackage{graphicx}
\usepackage{epsfig}
\usepackage{bm}
\usepackage[T1]{fontenc}
\usepackage[latin9]{inputenc}
\usepackage{amssymb}
\usepackage{float}
\usepackage{amsmath}
\usepackage{dcolumn}
\usepackage{cancel}
\usepackage[colorlinks]{hyperref}
\usepackage[usenames,dvipsnames]{color}
\hypersetup{
     breaklinks=true,
    pdfstartview={FitH},    % fits the width of the page to the window
    colorlinks=true,       % false: boxed links; true: colored links
    linkcolor=blue,          % color of internal links
    citecolor=red,        % color of links to bibliography
    filecolor=magenta,      % color of file links
    urlcolor=blue,           % color of external links
    anchorcolor=green,      % Color for anchor text
    linktocpage=true
}
\usepackage{amsmath}

\DeclareMathOperator{\arcsinh}{ash}

\newcommand{\be}{\begin{equation}}
\newcommand{\ee}{\end{equation}}

\begin{document}

\title{$P-v$ criticalities, phase transitions and geometrothermodynamics of charged AdS black holes from Kaniadakis statistics}

\author{G.~G.~Luciano}
\email{giuseppegaetano.luciano@udl.cat}
\affiliation{Applied Physics Section of Environmental Science Department, Escola Polit\`ecnica Superior, Universitat de Lleida, Av. Jaume
II, 69, 25001 Lleida, Spain}

\author{E. N. Saridakis}
\email{msaridak@noa.gr}
 \affiliation{National Observatory of Athens, Lofos Nymfon, 11852 Athens, 
Greece}
\affiliation{CAS Key Laboratory for Researches in Galaxies and Cosmology, 
Department of Astronomy, University of Science and Technology of China, Hefei, 
Anhui 230026, P.R. China}
 \affiliation{Departamento de Matem\'{a}ticas, Universidad Cat\'{o}lica del 
Norte, 
Avda.
Angamos 0610, Casilla 1280 Antofagasta, Chile}

\begin{abstract}
Boltzmann entropy-based thermodynamics of charged  anti-de Sitter (AdS) black 
holes has been shown to exhibit physically interesting features, such as $P-V$ 
criticalities and van der Waals-like phase transitions. In this work we extend 
the study of these critical phenomena to Kaniadakis theory, which is a 
non-extensive generalization of the classical statistical mechanics 
incorporating relativity. By applying the typical framework of condensed-matter 
physics, we analyze the impact of Kaniadakis entropy onto the equation of state, 
the Gibbs free energy and the critical exponents of AdS black holes in the 
extended phase space. Additionally, we investigate the underlying 
micro-structure of black holes in Ruppeiner geometry, which reveals appreciable 
deviations of the nature of the particle interactions from the standard 
behavior. Our analysis opens up new perspectives on the understanding of black 
hole thermodynamics in a relativistic statistical framework, highlighting the 
role of non-extensive corrections in the AdS black holes/van der Waals fluids 
dual picture. 
 \end{abstract}

 \maketitle

\section{Introduction}
\label{Intro} 

It is widely believed that black hole (BH) physics provides a promising arena to 
explore the quantum nature of gravity. After the pioneering discovery that BHs 
behave as thermodynamic systems~\cite{Hawking:1975vcx,Bek1}, the study of their 
properties has received a boost of new interest, and insights emerged toward the 
unification of general relativity, quantum theory and statistical 
physics~\cite{Bek2,Hawking:1982dh}. Yet, although BH dynamics can be fully 
described by a small number of classical parameters (namely  mass, angular 
momentum and charge - \emph{no hair theorem}),  
the microscopic degrees of freedom responsible for the
thermal behavior of BHs have not yet been adequately identified~\cite{Wald:1999vt}. 

The development of thermodynamic geometry (geometrothermodynamics) 
based on Weinhold~\cite{Wein1} and Ruppeiner~\cite{Rupp1,Rupp2} formalisms is an 
effort to extract, phenomenologically or qualitatively, the microscopic interaction information of a given system from the axioms of 
thermodynamics. 
The core idea is that, in ordinary thermodynamic systems, the curvature of 
Weinhold and Ruppeiner metrics is related to the nature of interactions among 
the underlying particles. For systems
where the micro-structures interact attractively, the curvature scalar carries a 
negative sign, whereas it is positive for predominantly repulsive forces. 
Moreover, the metric is flat 
for non-interacting systems - such as the ideal gas - or systems where interactions are perfectly balanced. 
This scheme has been tested for a wide number 
of statistical physical models \cite{Rupp2}. Interestingly enough, recent 
studies have revealed that it is feasible for BHs 
too~\cite{Cai:1998ep,Wei:2015iwa,Wei:2019uqg,Guo:2019oad,Xu:2020gud,
Ghosh:2020kba,Xu:2020ftx,Prom,Dehghani:2023yph,Santos:2023eqp},  
providing an empirical tool to access the microstructure of BHs
from their macroscopic knowledge, despite the absence of a quantum gravitational theory.

Black holes in anti-de Sitter (AdS) spacetimes have been thoroughly studied  in 
the last decades due to their applications in holography~\cite{Witt}. 
The observation that asymptotically AdS BHs can be described by dual thermal 
field theory  has motivated a parallel study with condensed matter systems. This has led to the discovery of 
first order phase transitions in BHs~\cite{Chamb1,Chamb2,Niu} that resemble in many 
aspects  the liquid-gas change of phase of van der Waals 
fluids~\cite{Sahay1,Sahay2,MannPT}. A constitutive ingredient of this picture is 
the (negative) cosmological constant $\Lambda$, which is identified as pressure and
included in the first law of BH thermodynamics alongside its 
conjugate quantity - the thermodynamic volume~\cite{Kastor,Dolan}. 
The ensuing \emph{extended phase
space} allows to formulate the $P=P(V,T)$ equation of state and study the 
critical  behavior of AdS BHs~\cite{Dolan,Dolan2}. 

A subtle concept in BH physics is thermodynamic entropy. According to  the 
holographic principle~\cite{tHooft:1993dmi,Susskind:1994vu}
BHs could store information at the event horizon like holograms.  In the 
standard Boltzmann-Gibbs statistics, this behavior is encoded by the 
Bekenstein-Hawking area law, which states that BH entropy scales like the 
surface area
\be
\label{SBH1}
S_{BH}=\frac{A_{bh}}{A_0}\,, 
\ee
where $A_0=4$ is the Planck area\footnote{We adopt  
geometric units $\hslash=c=k_B=G=1$.}. Clearly, this is an unconventional 
scaling. Indeed,
if BHs are physically identified with their event horizon surface, then they can 
be regarded as genuine $(2+1)$-dimensional systems and $S_{BH}$ is with the 
correct (extensive) thermodynamic entropy. However, if BHs are to be considered 
as $(3+1)$-dimensional objects (as arguably more natural in a 
$(3+1)$-dimensional description of the spacetime background), then the area 
scaling would violate thermodynamic extensivity. Thus, Boltzmann-Gibbs theory 
may not be the appropriate framework for studying the thermodynamics of BHs, and 
a generalized non-additive  entropy notion~\cite{Cirto} or a quasi-homogeneous 
black hole thermodynamics~\cite{Quevedo} could be needed for such non-standard 
systems.

To better  understand the intimate nature of BH entropy, several extensions of 
Boltzmann-Gibbs statistics have been considered in literature, motivated by 
either gravitational considerations (Tsallis~\cite{Tsallis,Cirto},  
Barrow~\cite{Barrow} and more generalized~\cite{Nojiri} entropies) or 
information theory (R\'enyi~\cite{Reny} and Sharma-Mittal~\cite{Sharma} 
entropies). Predictions of these models have been tested in
cosmology~\cite{SarBar,SarTs,SheTs,LucPRD,LucPLB,ShePert} and quantum physics~\cite{Shaba,LucTsGup,Luciano:2021mto,JizbaLamb}. Recently, a non-extensive
generalization  inspired by the symmetries of the relativistic  Lorentz group 
has been proposed by Kaniadakis~\cite{Kania0,Kania1,Kania2,Scarf1,Scarf2} based on the 
modified entropy
\be
\label{KE}
S_\kappa\,=\,-\sum_{i}n_i \ln_\kappa n_i\,,
\ee
where the $\kappa$-deformed logarithm is defined by
\be
\ln_\kappa x\,\equiv\, \frac{x^\kappa-x^{-\kappa}}{2\kappa}\,.
\ee
The generalized Boltzmann factor for the $i$-th microstate is
\be
\label{md}
n_i\,=\,\alpha \exp_\kappa\left[-\beta\left(E_i-\mu\right)\right],
\ee
where
\begin{eqnarray}
\exp_\kappa(x)&\equiv&\left(\sqrt{1+\kappa^2\,x^2}\,+\,\kappa\,x\right)^{1/\kappa}\,,\\[2mm]
\alpha&=&\left[(1-\kappa)/(1+\kappa)\right]^{1/2\kappa}\,,\\[2mm]
1/\beta&=&\sqrt{1-\kappa^2}\,\hspace{0.2mm}T\,,
\end{eqnarray}
with $T$ and $\mu$ being the temperature and chemical potential of the  system, 
respectively. 

Deviations from 
Boltzmann-Gibbs statistics are quantified by the
dimensionless parameter $-1<\kappa<1$. The classical framework
is, however, recovered in the $\kappa\rightarrow0$ limit. 
Besides theoretical arguments, we emphasize that phenomenological evidences  for 
Kaniadakis statistics 
come from the high-quality agreement between the modified 
distribution~\eqref{md} and  the observed power-law tailed spectrum of cosmic 
rays~\cite{Kania1}. 

One can show that, for the case of BHs,  Kaniadakis 
entropy~\eqref{KE} can be cast 
as~\cite{Luciano:2022eio,Morad,LympKan,Drepanou:2021jiv,She}
\be
\label{KEn}
S_\kappa\,=\,\frac{1}{\kappa}\sinh\left(\kappa\,S_{BH}\right)\,.
\ee
We mention here that, since the above expression is an even function  of 
$\kappa$, i.e. $S_\kappa=S_{-\kappa}$, in the following we shall restrict to the 
$\kappa\ge0$ domain. 

Kaniadakis entropy in the form~\eqref{KEn} has been mostly used
for holographic applications and, in particular, to infer corrections brought 
about in the Friedmann equations~\cite{Morad,LympKan, 
Drepanou:2021jiv,She,Hernandez-Almada:2021aiw,Hernandez-Almada:2021rjs,
Lambiase:2023ryq} (see also~\cite{Luciano:2022eio} for a recent review). In the 
light of the gravity-thermodynamic conjecture, preliminary studies in BH 
thermodynamics
have been considered in~\cite{Cimi} by computing 
the $\kappa$-deformed temperature and heat capacity in the
context of generalized Heisenberg relations~\cite{Kempf,Buoninf}.
Nevertheless, to the best of our knowledge, a dedicated analysis 
of BH geometrothermodynamics and critical phenomena
in Kaniadakis statistics has not yet been conducted. 

Starting from the above premises, in this work we 
address the thermodynamics of AdS BHs from the Kaniadakis entropy  perspective. 
We investigate the impact of Eq.~\eqref{KEn} on small-large BH phase transitions 
and critical exponents by exploiting the language of condensed matter physics. 
In this sense, the main effort here is to lay the foundation towards formulating 
BH thermodynamics in a fully relativistic statistical context. We then examine 
the underlying microstructure of BHs in Ruppeiner geometry, which 
reveals predominantly repulsive intermolecular forces. In line with the 
discussion  of~\cite{NRGS1,NRGS4,NRGS5,NRGS6,NRGS8,Inl1,Inl2}, our analysis shows that 
the development of BH thermodynamics based on a non-extensive  entropy notion involves a consistent redefinition of all other thermodynamic 
quantities, including the Hawking temperature and thermodynamic energy. 

The structure of the work is as follows:
for later comparison with physics of BHs,
the next Section is devoted to review phase transitions
and critical phenomena of van der Waals fluids. 
In Sec.~\ref{ChargAdS} we study thermodynamics of
charged AdS BHs in Kaniadakis statistics, while Sec.~\ref{Geom}
concerns geometrothermodynamic analysis. 
Conclusions and perspectives are finally discussed in Sec.~\ref{Disc}.

\section{$P-V$ criticality of van der Waals fluids}
\label{PVvdW}

Van der Waals model provides an effective description of
real interacting fluids and 
liquid-gas phase transitions. The characteristic equation is
\be 
\label{vdwequation}
\left(P+\frac{a}{v^2}\right)\left(v-b\right) = T\,, 
\ee 
where $v=V/N$, $N$, $V$, $P$ and $T$ 
denote the specific volume, number of constituents, global volume,  pressure and 
temperature of the van der Waals system, respectively.
The positive constant $a$ and $b$ 
quantify the attraction and finite size of the molecules in the fluid. 

\begin{figure}[t]
\begin{center}
\hspace{-3mm}\includegraphics[width=8cm]{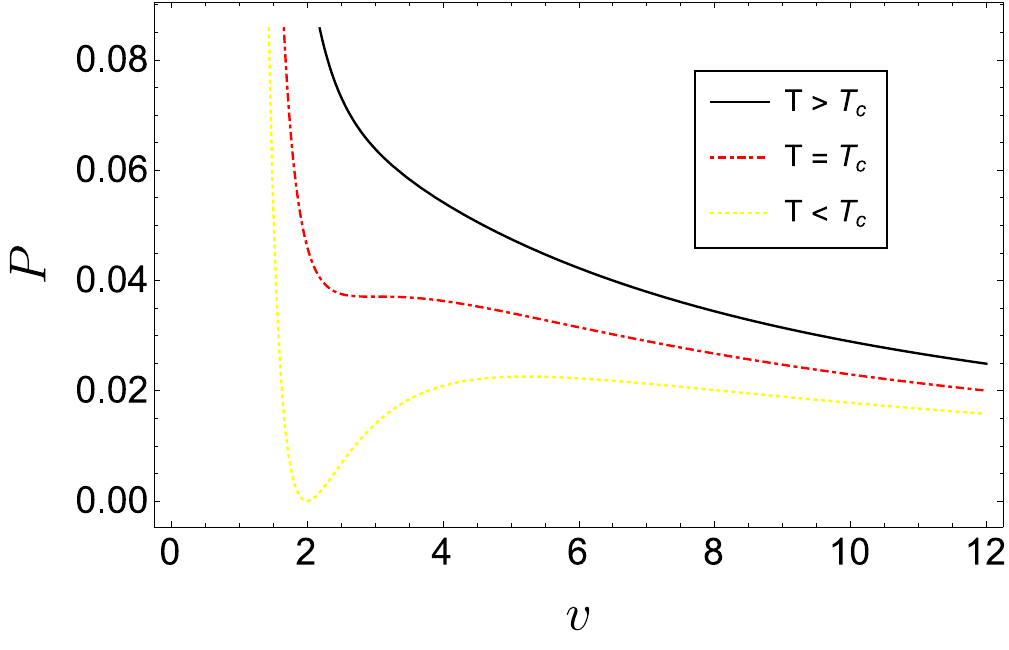}
\caption{{\it{$P-v$ diagram of van der Waals fluids.  The red dot-dashed line 
indicates the critical isotherm at $T=T_c$. We have set $a=b=1$ (online 
colors).}}}
\label{Fig1}
\end{center}
\end{figure}

The qualitative behavior of $P-V$ isotherms is displayed in
Fig.~\ref{Fig1}. It can be seen that the \emph{critical point}
of the liquid-gas phase transition occurs when $P(v)$ has an inflection point, 
which is obtained by imposing 
\be
\label{C1}
\left(\frac{\partial P}{\partial v}\right)_T=
\left(\frac{\partial^2 P}{\partial v^2}\right)_T=0\,.
\ee 
In this way, we obtain
\begin{equation}
\label{cVan}
v_{c}=3b\,, \quad \,
T_{c}=\frac{8a}{27b}\,, \quad \,
P_{c}=\frac{a}{27b^2}\,,
\end{equation}
for the critical volume, temperature and pressure, respectively.
It is immediate to check that  
\be
\label{pcvctc}
P_c v_c/T_c=3/8\,,
\ee 
which is a universal number predicted for all fluids (independently of the 
constants $a$ and $b$).

\begin{figure}[t]
\begin{center}
\hspace{-3mm}\includegraphics[width=8cm]{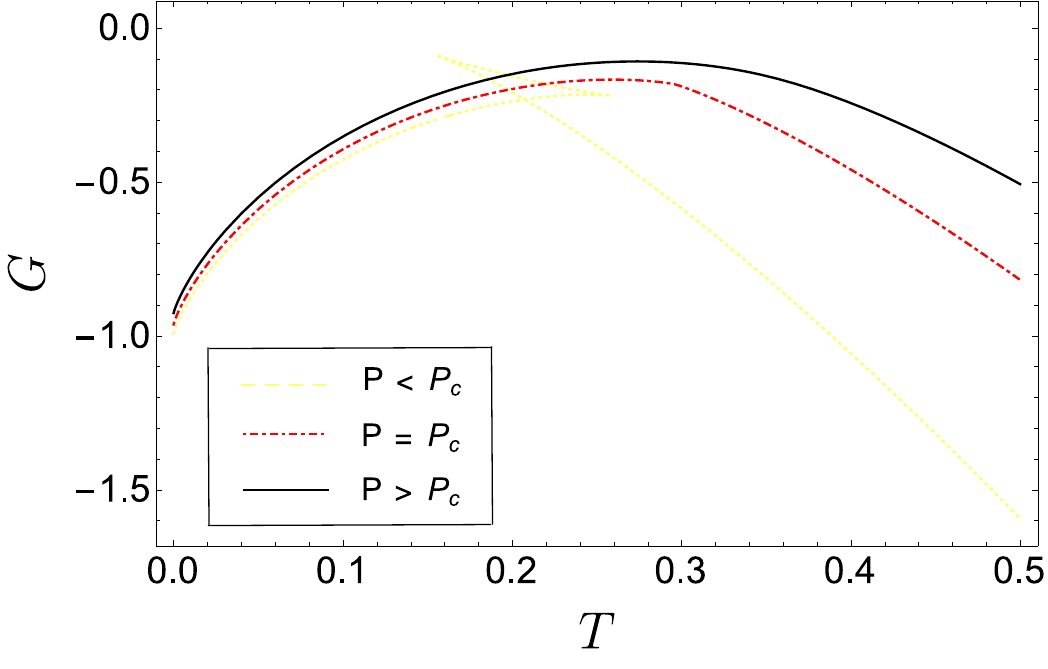}
\caption{{\it{ Gibbs free energy $G$ versus temperature $T$ for various 
pressures $P$, for van der Waals fluids.
The red dot-dashed line indicates the critical isobar at $P=P_c$ (online 
colors).}}}
\label{Fig2}
\end{center}
\end{figure}

To gain more insights on the phase transitions of van der Waals fluids, let us introduce
the (specific) Gibbs free energy, $G = G(P, T)$. For fixed $N$, 
this is given by~\cite{MannPT}
\be
G(T,P)=-T\left\{1+\log\left[\frac{\left(v-b\right)T^{\frac{3}{2}}}{\lambda}
\right]\right\}-\frac{a}{v}+P
v\,, 
\ee 
where $v$ is to be understood as a function of pressure
and temperature through Eq.~\eqref{vdwequation}, 
while  $\lambda$ is a (dimensional) 
specific constant of the gas. The behavior of 
$G$ versus $T$ is shown in Fig.~\ref{Fig2} for different $P$. 
Below the critical pressure (yellow dotted curve), it
exhibits the ``swallow-tail'' shape characteristic
of first order phase transitions from liquid to gas. Such a feature disappears for $P>P_c$ (black solid line). 

The behavior of the physical variables near the critical point
is quantitatively described by the critical exponents. 
Following~\cite{MannPT}, we introduce
\be
\label{TildeQ} 
t=\frac{T-T_c}{T_c}=\tau-1\,,\, \quad\,\, 
\phi=\frac{v-v_c}{v_c}=\nu-1\,. 
\ee 
The basic critical exponents $\alpha,\beta,\gamma$ and $\delta$ 
are then defined as follows (for computational details, see~\cite{Gold}): 
\begin{itemize}

\item [-] $\alpha$ governs the dynamics of the specific heat at constant volume 
$C_v$ according to $C_v=T\left(\frac{\partial S}{\partial T}\right)_v \propto 
|t|^{-\alpha}$\,. By explicit computation, 
one sees that $C_v$ does not depend on $t$, which implies
$\alpha=0$.

\item[-] $\beta$ describes the behavior of the order parameter $\eta= v_g-v_l$ 
for a given isotherm as $\eta\propto |t|^{\beta}$\,, where
$v_{g,l}$ denote the volume of the gas and liquid phases, respectively. From the 
equation of
corresponding states for van der Waals fluids 
and the Maxwell's equal area law, it follows that $\beta=1/2$.

\item[-] $\gamma$ measures the isothermal compressibility $\kappa_T$ of the fluid in compliance with $\kappa_T=-\frac{1}{v}\left(\frac{\partial v}{\partial 
P}\right)_T\propto |t|^{-\gamma}$\,. By using again the equation of
corresponding states, one finds $\gamma=1$.

\item[-] $\delta$ controls the difference $|P-P_c|$ on the critical isotherm 
$T=T_c$ according to $|P-P_c|\propto|v-v_c|^{\delta}$. The study
of the shape of the critical isotherm gives $\delta=3$. 
\end{itemize}
 
The above considerations provide the basics of our next analysis.
Specifically, we elaborate on the correspondence
between phase transitions of BHs and van der Waals fluids
within Kaniadakis framework, with focus on the $\kappa$-deformed
analogues of Eq.~\eqref{vdwequation}-\eqref{TildeQ}.

\section{Kaniadakis thermodynamics of charged AdS black holes}
\label{ChargAdS}

The general static and spherically symmetric metric that describes 
$(3+1)$-dimensional charged AdS BHs in the Schwarzschild coordinates 
$(t,r,\theta,\phi)$ is given by~\cite{Re} 
\be
\label{metric}
ds^2=-f(r)d t^2+ f( r)^{-1}d r^2+r^2d\Omega^2\,,
\ee
where $d\Omega^2=d\theta^2+\sin^2\theta d\phi^2$ is the angular part of the 
metric on the two-sphere. Here, we have defined\footnote{Following the study 
of~\cite{Saridakis:2020lrg,Rani:2022xza,Jawad:2022,Luciano:2023fyr,
Basilakos:2023kvk} in Tsallis and Barrow frameworks, here we assume that 
Kaniadakis model only modifies BH entropy, while leaving the field equations of 
the theory unaffected. A comprehensive analysis of BH thermodynamics involving 
an ab initio derivation of a lagrangian driven by the Kaniadakis entropic index 
is reserved for the future.}
\be
f(r)=1-\frac{2 M}{ r}+\frac{Q^2}{r^2}+\frac{ r^2}{l^2}\,,
\ee
where $M,Q$ are the mass and electric charge of the BH, respectively, 
while $l$ is the AdS radius related to
the (negative) cosmological constant by 
\be
\Lambda=-\frac{3}{l^2}\,.
\ee

In our setup, the parameter $M$ shall be
associated with the enthalpy of the BH conceived as a thermodynamic
system. Clearly, for $Q=0$ and $l\gg r$, Eq.~\eqref{metric}
reduces to the well-known  Schwarzschild metric. 
Additionally, the event horizon $r_+$ of the geometry~\eqref{metric} 
corresponds to the largest root of $f(r)=0$. One can use this
solution to express the BH mass as 
\be
\label{mass}
M(r_+)=\frac{r_+}{2} + \frac{Q^2}{2r_+}+\frac{r^3_+}{2l^2}\,.
\ee

On the other hand, the surface area $A_{bh}$ of the BH horizon 
reads
\be
A_{bh}=\int_{r=r_+}\sqrt{g_{\theta\theta}\,g_{\phi\phi}}d\theta d\phi =4\pi 
r_+^2\,.
\ee
Accordingly, the Boltzmann-Gibbs-based Bekenstein-Hawking entropy
obeys the area law
\be
\label{SBH}
S_{BH}=\frac{A_{bh}}{4}=\pi r_+^2\,.
\ee

As discussed in the Introduction, 
due to the area scaling of BH entropy, arguments
from multiple perspectives suggest that Boltzmann-Gibbs
statistics  may not be the
appropriate context for studying the thermodynamics of BHs.
In particular, in a relativistic scenario Eq.~\eqref{SBH} is expected to be generalized to Kaniadakis entropy~\eqref{KEn}, which we rewrite here by dropping for simplicity the index $\kappa$
\begin{equation}
S=\frac{1}{\kappa}\sinh\left(\kappa\,S_{BH}\right)\,.
\label{Kenbis}
\end{equation}

Two comments are in order: first, 
we observe that $S$ is a monotonically increasing function of $S_{BH}$ and, 
thus, of the horizon radius $r_+$. Furthermore, in order to provide analytical 
solutions, it proves sometimes convenient to perform Taylor expansions for small 
$\kappa$~\cite{LympKan,LucBar}.  
This assumption is substantiated by the best agreement between theoretical predictions and 
phenomenological implications of Kaniadakis model, which is obtained for 
$\kappa=0.2165$ in the physics of cosmic rays~\cite{Kania1}. On the other hand, 
observational Cosmology constrains Kaniadakis parameter around 
zero~\cite{Luciano:2022eio}, while the interval $0<\kappa<1$ has been 
considered 
in connection with the Bekenstein bound conjecture in Schwarzschild 
BHs~\cite{Abreu:2022pil}. 
In what follows, we shall retain the exact expression of $S$ as far as possible, 
resorting to the small $\kappa$-approximation when strictly necessary.

The thermodynamic picture of AdS BHs is completed by the introduction of an extended phase space, where 
the pressure is identified with the cosmological constant
and the thermodynamic volume with its conjugate quantity, i.e.
\begin{eqnarray}
\label{pressure}
P&=&-\frac{\Lambda}{8\pi}=\frac{3}{8\pi l^2}\,,\\[2mm]
V&=&\left(\frac{\partial M}{\partial P}\right)_{S,Q}
=\frac{4}{3}\pi r_+^3\,,
\label{Vol}
\end{eqnarray}
respectively. Equipped with these new definitions, it is easy to check that BHs still obey the first law 
of thermodynamics~\cite{MannPT}
\be
\label{FLT}
dM=TdS +\varphi dQ+VdP\,,
\ee
and Smarr relation
\be
\label{Smarr}
M = 2\left(TS-VP\right)+\Phi Q\,,
\ee
where 
\be
T=\left(\frac{\partial M}{\partial S}\right)_{P,Q}\,,\quad\, 
\Phi=\left(\frac{\partial M}{\partial Q}\right)_{S,P}\,,
\ee 
are the temperature and electric potential, respectively. 

\begin{figure}[t]
\begin{center}
\hspace{-3mm}\includegraphics[width=7.6cm]{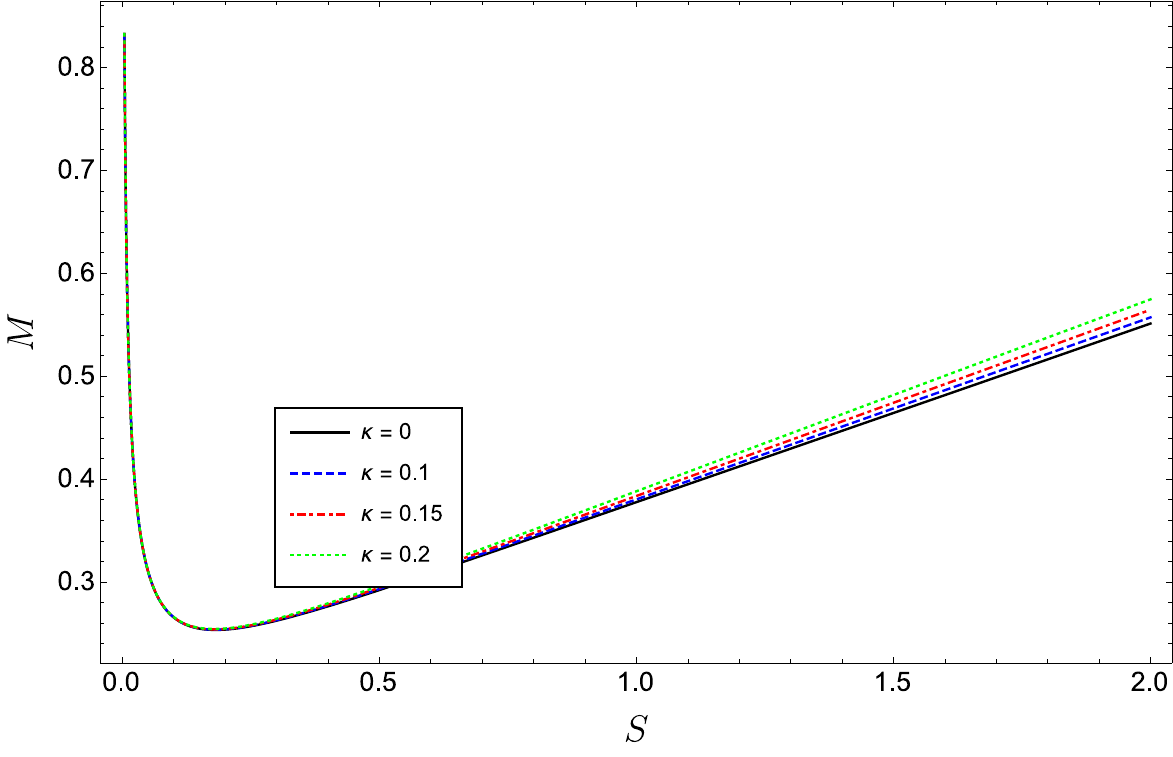}
\\[0.5cm]
\hspace{-3mm}\includegraphics[width=7.6cm]{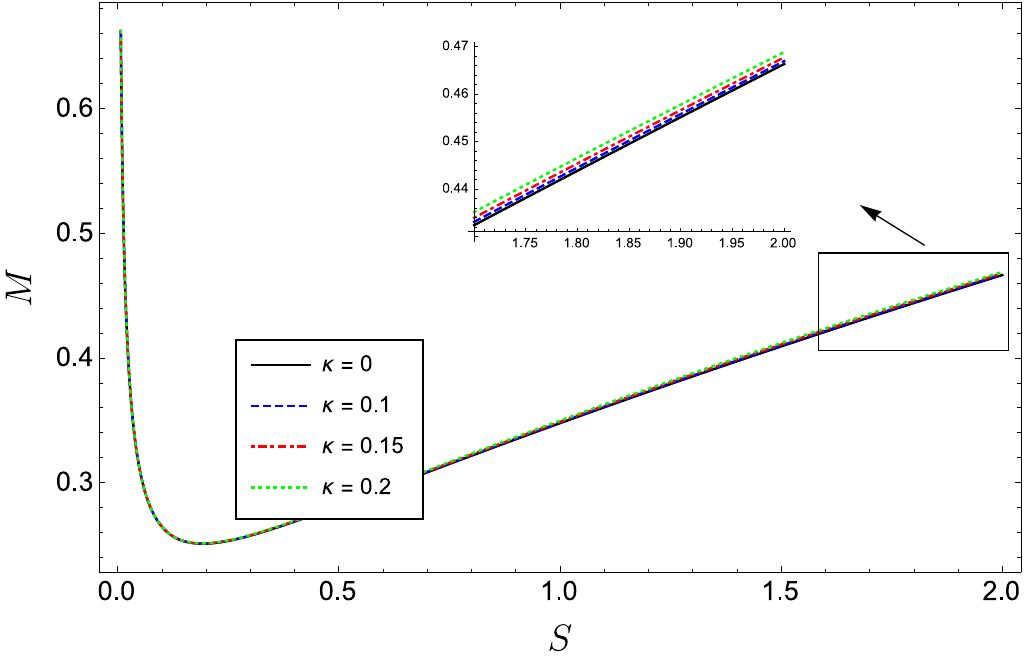}
\\[0.5cm]
\hspace{-3mm}\includegraphics[width=7.6cm]{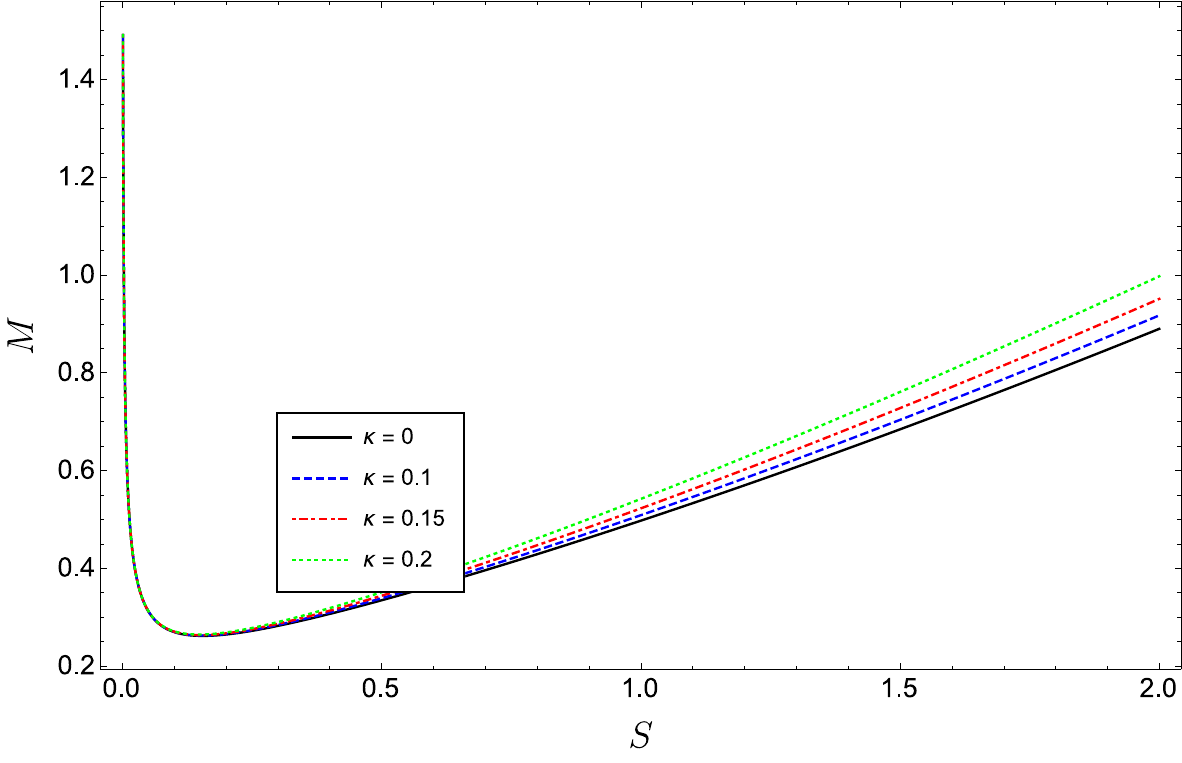}
\caption{{\it{The mass parameter  $M$ versus the entropy $S$ for various 
values of $\kappa$, and for  $l=l_c$ fixed through the critical 
condition~\eqref{Pc} (upper panel), $l=2l_c$ (middle panel) and $l=0.5l_c$ 
(lower panel) (online colors).}}}
\label{Fig3}
\end{center}
\end{figure}

Using the standard thermodynamic 
machinery, we now have all the ingredients to
compute the necessary BH thermodynamic variables. Since 
BH phase transitions have been shown to occur in the canonical (fixed charge) 
ensemble~\cite{Chamb1,Chamb2}, 
we conduct our analysis in this framework. As a first step, we express the mass 
parameter~\eqref{mass} in terms of the Kaniadakis entropy, using the 
relation~\eqref{KEn}. This gives
\be
\label{M}
M(S)=\frac{\left(\pi l Q \kappa\right)^2+\pi l^2 \kappa \arcsinh\left(\kappa 
S\right)+\arcsinh^2\left(\kappa 
S\right)}{2\pi^{\frac{3}{2}}l^2\kappa^{\frac{3}{2}}\arcsinh^{\frac{1}{2}}
\left(\kappa S\right)}\,,
\ee
where we have introduced the shorthand notation 
\be
\arcsinh\left(x\right)\equiv \mathrm{arcsinh}\left(x\right)\,.
\ee
One can verify that  
the limit for $\kappa\rightarrow0$  of Eq.~\eqref{M} reproduces the standard 
expression of $M$ for charged AdS BHs (solid black lines in Fig.~\ref{Fig3}), 
namely
\be
M_{\kappa\rightarrow0}(S)=\frac{S^2+\pi l^2\left(\pi 
Q^2+S\right)}{2\pi^{\frac{3}{2}}l^2 S^{\frac{1}{2}}}\,.
\ee

The behavior of the $\kappa$-deformed mass~\eqref{M} versus $S$ is shown in 
Fig.~\ref{Fig3} for various $\kappa,l$ and fixed $Q=0.25$ as 
in~\cite{Soroushfar}. According to our previous considerations on the expected 
smallness of Kaniadakis exponent, we restrict to the range $\kappa\le0.2$. Note that in 
cosmological applications, $\kappa$ is found to be much closer to zero 
\cite{Hernandez-Almada:2021aiw,Hernandez-Almada:2021rjs}, however this does not 
need to be the case in 
BH applications (nevertheless, even in the present analysis one could use arbitrarily small values of the entropic parameter). As we can see, $M$ remains positive and  
shows an initially decreasing behavior (evaporation phase of the BH), followed by a later 
growth (absorption process). While leaving the initial stage of the evolution 
nearly unaffected, Kaniadakis entropy influences the final growth rate of $M$. 
In particular, the higher $\kappa$, the faster the growth, and vice-versa.
For the sake of comparison with recent literature, we emphasize that a similar 
result has been found in the context of quantum gravity-induced deformations of 
Boltzmann-Gibbs entropy~\cite{Jawad:2022} and in non-linear electrodynamics and 
the Einstein-massive gravity~\cite{Nam}. 

\begin{figure}[t]
\begin{center}
\includegraphics[width=7.6cm]{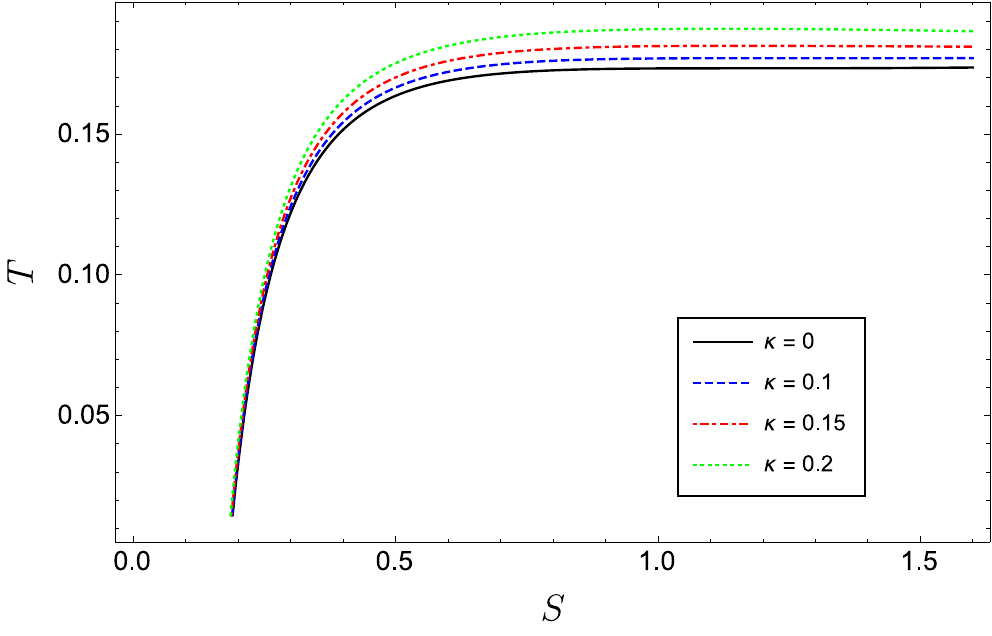}
\\[0.5cm]
\includegraphics[width=7.6cm]{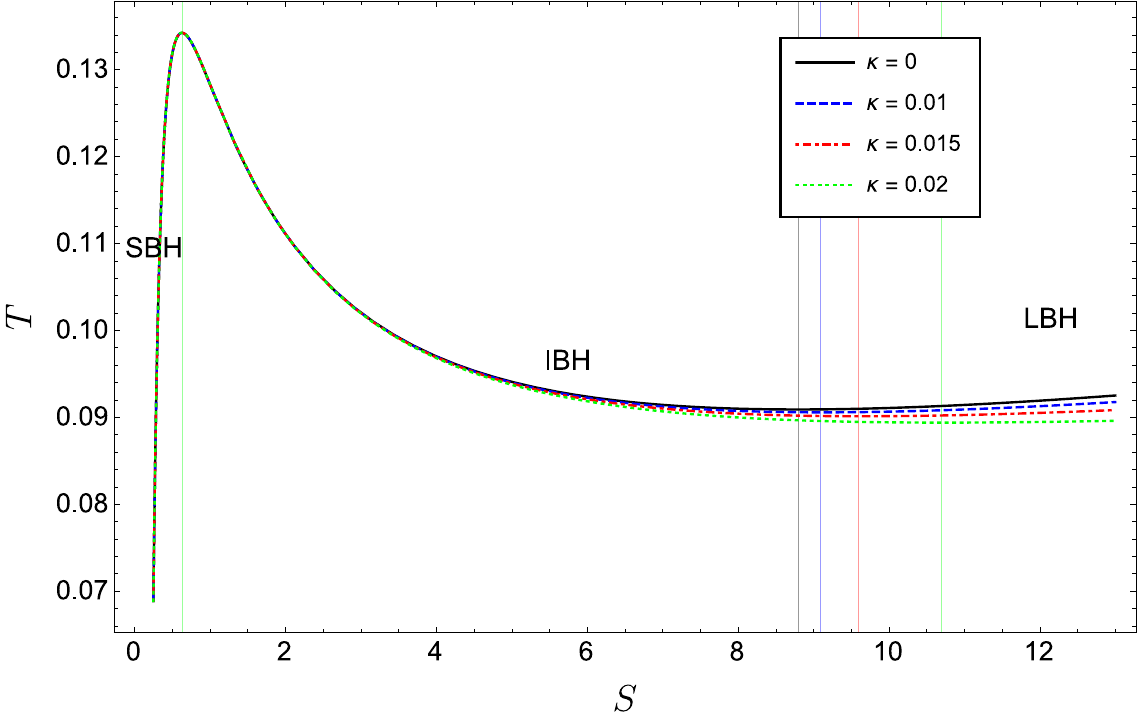}
\\[0.5cm]
\hspace{3mm}\includegraphics[width=7.6cm]{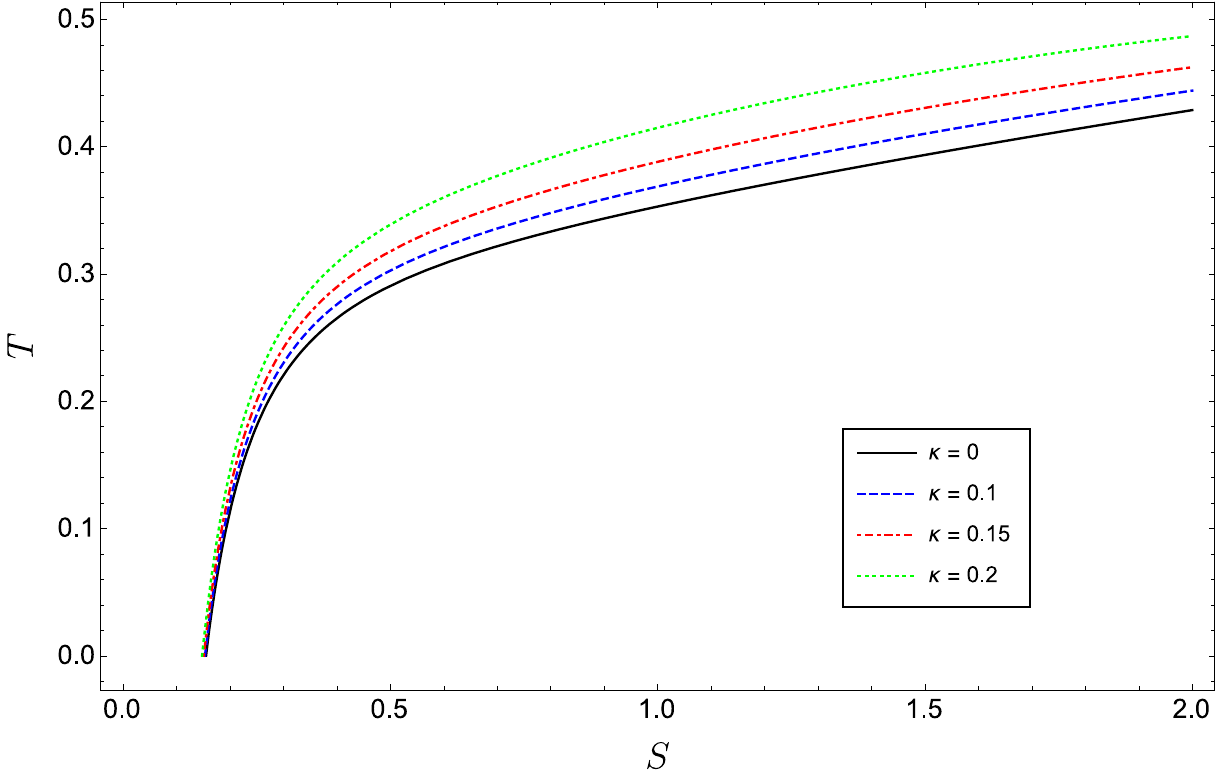}
\caption{{\it{The temperature  $T$ versus the entropy $S$ for various 
values of $\kappa$, and for $l=l_c$ fixed through the critical 
condition~\eqref{Pc} (upper panel), $l=2l_c$ (middle panel) and $l=0.5l_c$ 
(lower panel). In order to reveal all   features of the $T-S$ 
diagrams, in the middle panel we have additionally considered smaller values of 
$\kappa$. The vertical lines in the middle panel separate  Region II 
- Intermediate Black Hole (IBH) - from Region I - Small Black Hole (SBH) - and Region III - Large Black Hole (LBH) - see text (online colors).
}}}
\label{Fig4}
\end{center}
\end{figure}

The usage of Eq.~\eqref{M} along with the first law of 
thermodynamics~\eqref{FLT} allows us to derive the $\kappa$-temperature of the 
thermal radiation emitted by BHs as
\begin{eqnarray}
\label{T}
T(S)&=&\left(\frac{\partial M}{\partial S}\right)_{P}\\[2mm]
\nonumber
&=&\frac{-\left(\pi l Q \kappa\right)^2+\pi l^2 \kappa\arcsinh\left(\kappa 
S\right)+3\arcsinh^2\left(\kappa 
S\right)}{4\pi^\frac{3}{2}l^2\left[\left(1+\kappa^2S^2\right)\kappa 
\arcsinh^3\left(\kappa S\right)\right]^{\frac{1}{2}}}\,,
\end{eqnarray}
which, in the limit of $\kappa\rightarrow0$, still recovers the usual result (solid black line in 
Fig.~\ref{Fig4})
\be
T_{\kappa\rightarrow0}(S)=\frac{3S^2+\pi l^2\left(S-\pi 
Q^2\right)}{4\pi^{\frac{3}{2}}l^2 S^{\frac{3}{2}}}\,.
\ee
We further notice that the above relation has the well-defined Schwarzschild limit $T=1/(4\pi r)$ for $Q\rightarrow0$ and $l\rightarrow\infty$. 

Equation~\eqref{T} is plotted as a function of $S$ in Fig.~\ref{Fig4} for 
various $\kappa,l$ and fixed $Q$ as before. 
The points where the slope of the $T-S$ graphs vanishes are of special interest, 
as they signal a potentially critical behavior of BHs (see the next section for 
more quantitative discussion).  From the upper and lower panel of 
Fig.~\ref{Fig4}, we observe that $T$ increases monotonically and has one (upper panel) or no (lower panel) 
stationary point, depending on the value of $l$ (or, equivalently, 
of $P$). 

On the other hand, the middle panel 
shows that $T$ increases for small (Region I - Small Black Hole (SBH)) and large 
(Region III - Large Black Hole (LBH)) values of the entropy, while it decreases 
in the intermediate domain (Region II - Intermediate Black Hole (IBH)). These 
regions are separated by two stationary points, 
which have been marked by vertical lines for better visual clarity.
Effects of Kaniadakis entropy manifest through a variation of the width of the 
IBH region, with higher $\kappa$ yielding a larger IBH domain and vice-versa. 
Below, we shall see this  behavior of $T-S$ graph  is peculiar to a first-order 
phase transition
between SBH and LBH, which resembles in many aspects the liquid-gas change of 
phase of van der Waals fluids. In this picture, the larger amplitude of the IBH 
domain for higher $\kappa$
can be understood in terms of the non-extensive character of Kaniadakis entropy 
by looking at the BH structure on a molecular level (see Sec.~\ref{Geom}). For 
higher
$\kappa$, indeed, the repulsive forces
among BH microstructures 
tend to be weaker, at least in the first stage of BH evolution, which implies a 
slowed small-to-large phase transition of BH. 
To carry on the similarity with van der Waals-like systems (where, however, 
interparticle forces are mostly attractive), one can think of a Kanidakis BH as 
a fluid with stronger attraction for larger $\kappa$. In this case, more heat is necessary 
to the internal molecules to overcome these attractive interactions, which results in a delayed liquid-gas change of phase.

Finally, regardless of the value of $l$, the condition $T(S_0)=0$
gives the \emph{physical limitation point of BHs}. Indeed,
for $S<S_0$ the temperature becomes negative, which means this region is physically inaccessible. 

Before moving on, we remark that the employment of generalized (non-extensive) entropies like that in Eq.~\eqref{KEn} leads to multiplicity in the temperature value of BHs. In~\cite{NRGS8} it has been
observed that three different scenarios may occur, depending on the assumed energy and temperature definitions. In compliance with~\cite{Rani:2022xza,Jawad:2022}, here we are considering the energy definition of GR, the first law of thermodynamics and the thermodynamic temperature definition as fundamental. An alternative viewpoint has been adopted in~\cite{NRGS8}, based on the assumption that the Hawking temperature must be kept unaffected. It is interesting to explore whether, and if so, how the present results get modified in such  a complementary approach. Investigation along this direction is left for future work.

\subsection{Heat capacity and critical point}
\label{HC}

\begin{figure}[t]
\begin{center}
\hspace{-6mm}\includegraphics[width=8cm]{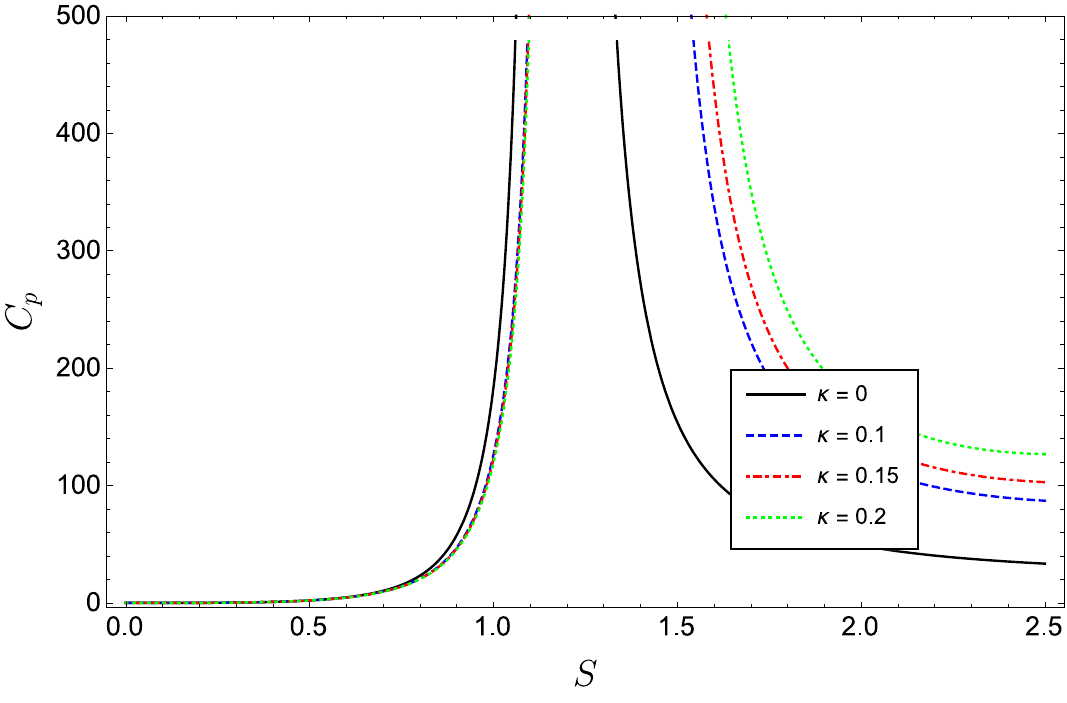}
\\[0.5cm]
\hspace{-6mm}\includegraphics[width=8cm]{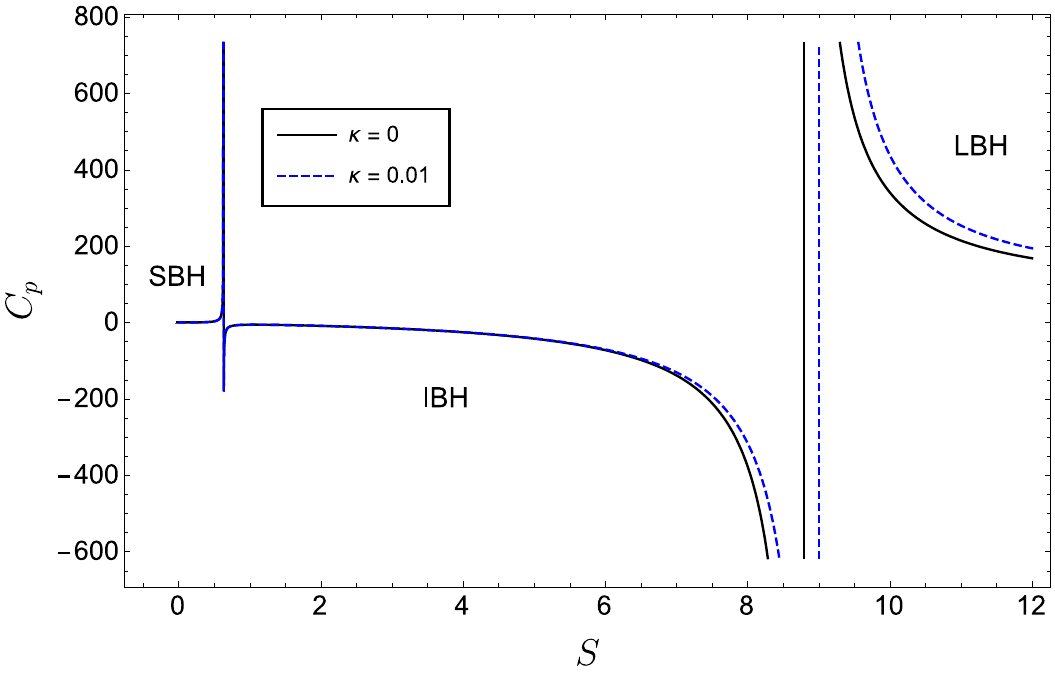}
\\[0.5cm]
\hspace{-3mm}\includegraphics[width=8cm]{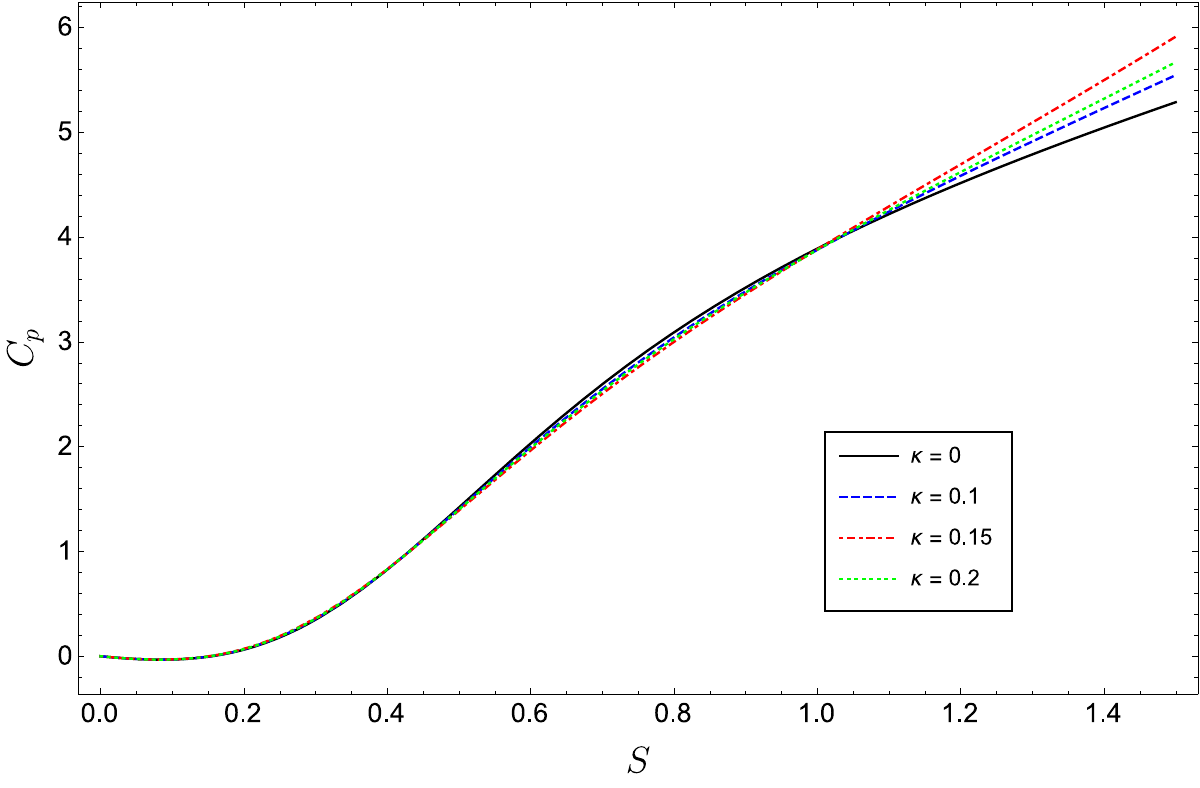}
\caption{{\it{The heat capacity at constant pressure  $C_p$ versus the entropy 
$S$ for various $\kappa$ values,  and for $l=l_c$ fixed through the critical 
condition~\eqref{Pc} (upper panel), $l=2l_c$ (middle panel) and $l=0.5l_c$ 
(lower panel). For visual clarity, we have only considered two values of 
$\kappa$ in the middle panel (online colors).}}}
\label{Fig5}
\end{center}
\end{figure}

With the help of the temperature~\eqref{T}, one can obtain the heat capacity
at constant pressure as
\begin{widetext}
\begin{eqnarray}
\nonumber
C_p(S)=T\left(\frac{\partial S}{\partial T}\right)_{P}&=&
-\frac{2}{\kappa}\left(1+\kappa^2S^2\right)^{\frac{3}{2}}\arcsinh\left(\kappa 
S\right)\left[\left(\pi l Q\kappa\right)^2-\arcsinh\left(\kappa 
S\right)\left(\pi l^2\kappa+3\arcsinh\left(\kappa S\right)\right)\right]\\[2mm]
\nonumber
&&\times\bigg\{3\left(\pi l 
Q\kappa\right)^2\left(1+\kappa^2S^2\right)-\arcsinh\left(\kappa S\right)
\Big\{\pi l^2\kappa \left[1+\kappa^2 S\left(S-2\pi Q^2(1+\kappa^2 
S^2)^\frac{1}{2}\right)
\right]\\[2mm]
&&+\arcsinh\left(\kappa S\right)
\left[-3+\kappa^2 S\left(-3S+2\pi 
l^2(1+\kappa^2S^2)^{\frac{1}{2}}\right)+6\kappa S\arcsinh\left(\kappa 
S\right)\left(1+\kappa^2 S^2\right)^\frac{1}{2}\right]
\Big\}
\bigg\}^{-1}\,,
\label{HeCa}
\end{eqnarray}
\end{widetext}
which for $\kappa\rightarrow0$ consistently reduces to (solid black line in 
Fig.~\ref{Fig5})
\be
C_{p,\kappa\rightarrow0}(S)=\frac{2S\left[3S^2+\pi l^2\left(S-\pi 
Q^2\right)\right]}{\pi l^2\left(3\pi Q^2-S\right)+3S^2}\,.
\ee

It is important to note that $C_p>0$
corresponds to local stability of BHs, 
while for $C_p<0$ even small perturbations
may cause BH disappearance. Also, discontinuities potentially indicate a 
critical behavior of BHs. 

Equation~\eqref{HeCa} is plotted 
as a function of $S$ in Fig.~\ref{Fig5} for various $\kappa,l$ and fixed $Q$ as 
before. In compliance with the discussion
below Eq.~\eqref{T}, it is observed that $C_p$ has one (upper panel), two 
(middle panel) or no (lower panel) discontinuity, 
depending on the value of $l$. The SBH, IBH and LBH regions are clearly 
distinguishable
from the middle panel (the vertical lines correspond to the two stationary 
points of of the $T-S$ graphs in the middle panel of Fig.~\ref{Fig4}). While SBH 
and LBH
are thermodynamically stable ($C_p>0$), 
IBH is unstable ($C_p<0$). As discussed, for instance, in~\cite{MannPT} this 
gives rise to a transition between the SBH and LBH ``phases''. As $l$ 
progressively decreases to a certain critical value (upper panel), the IMB range 
reduces to a point, which in turn
corresponds to the single stationary point of the $T-S$ graph in the upper panel 
of Fig.~\ref{Fig4}. By further decreasing $l$ (lower panel), $C_p$ is always 
continuous and keeps positive values. In this case, BHs are locally stable and 
do not undergo any transition (see also the corresponding $T-S$ graphs in the 
lower panel of Fig.~\ref{Fig4}). 

To better understand the origin of this critical value of $l$, we now derive
the analogue of the equation of state~\eqref{vdwequation} for BHs in the 
extended phase space. By using Eq.~\eqref{pressure} and~\eqref{T}, after some 
algebra we get
\be
\label{EoS}
P(r_+)=\cosh\left(\pi \kappa 
r_+^2\right)\hspace{0.4mm}\frac{T}{2r_+}+\frac{Q^2}{8\pi r_+^4}-\frac{1}{8\pi 
r_+^2}\,,
\ee
which is now straightforward to match with the $\kappa\rightarrow0$ 
limit~\cite{MannPT}
\be
P_{\kappa\rightarrow0}(r_+)=\frac{T}{2r_+}+\frac{Q^2}{8\pi r_+^4}-\frac{1}{8\pi 
r_+^2}\,.
\ee

To directly compare Eq.~\eqref{EoS} with Eq.~\eqref{vdwequation}, let us 
identify the horizon radius $r_+$
with the specific volume of van der Waals fluid as~\cite{MannPT}
\be
\label{defvrh}
v=2 r_+\,.
\ee
In this way, we obtain
\be
\label{EoS2}
P(v)=\cosh\left(\frac{\pi \kappa 
v^2}{4}\right)\hspace{0.4mm}\frac{T}{v}+\frac{2Q^2}{\pi v^4}-\frac{1}{2\pi 
v^2}\,.
\ee

As discussed for van der Waals fluids, 
the critical point of phase transitions  
can be derived from the conditions~\eqref{C1}. 
However, analytical expressions for the critical specific volume, temperature 
and pressure can only be obtained to the leading order in $\kappa$. In this approximation, we are allowed to write down
\begin{eqnarray}
\label{vc}
\hspace{-6mm}v_c&=&2\sqrt{6}Q+144\sqrt{6}\pi^2 
Q^5\kappa^2+\mathcal{O}(\kappa^3)\,,\\[2mm]
\label{Tc}
\hspace{-6mm}T_c&=&\frac{1}{3\sqrt{6}\pi Q}+3\sqrt{6}\pi 
Q^3\kappa^2+\mathcal{O}(\kappa^3)\,,\\[2mm]
\label{Pc}
\hspace{-6mm}P_c&=&\frac{1}{96\pi Q^2}+2\pi Q^2 
\kappa^2+\mathcal{O}(\kappa^3)\,,\quad l_c^2=\frac{3}{8\pi P_c}\,,
\end{eqnarray}
all reducing to the standard critical expressions for 
$\kappa\rightarrow0$~\cite{MannPT}.

From Eq.~\eqref{vc}, we can also infer the 
thermodynamic volume corresponding to the critical volume $v_c$, which is
\begin{eqnarray}
\nonumber
V_c&=&\left(\frac{\partial M}{\partial P}\right)_S\bigg|_{r_+=r_c}
=\frac{4}{3}\pi r_c^3\\[2mm]
&=&8\sqrt{6}\pi Q^3+1728\sqrt{6}\pi^3 Q^7\kappa^2+\mathcal{O}(\kappa^3)\,.
\end{eqnarray}

It is worth noting that $v_c\rightarrow0$, while $T_c,P_c\rightarrow\infty$ for 
$Q\rightarrow0$ regardless
of $\kappa$, which means 
that the critical transition
described above is characteristic of charged 
BHs also in Kaniadakis entropy model (on the other hand, in~\cite{GB} it has been found that AdS BHs in Gauss-Bonnet gravity undergo small-large transitions in the uncharged case too).

Interestingly enough, the critical parameters~\eqref{vc}-\eqref{Pc} satisfy the 
relation
\be
\frac{P_c v_c}{T_c}=\frac{3}{8}+\frac{315}{4}\pi^2 
Q^4\kappa^2+\mathcal{O}(\kappa^3)\,.
\ee
By comparison with Eq.~\eqref{pcvctc}, 
we see that Kaniadakis entropy-based BHs slightly deviate from pure van der Waals behavior, due to the $\kappa$-dependent correction. The latter 
behavior is, however, recovered for $\kappa\rightarrow0$.

Now, by introducing the re-scaled variables
\be
p=\frac{P}{P_c}\,,\quad \nu=\frac{v}{v_c}\,,\quad \tau=\frac{T}{T_c}\,,
\ee
the equation of state~\eqref{EoS2} can be rearranged as
\be
\label{lcs}
8\tau=3\nu\left(Ap+\frac{2 B}{\nu^2}\right)-\frac{D}{\nu^3}\,,
\ee
where 
\begin{eqnarray}
\nonumber
A&=& \frac{8 P_c v_c}{3 T_c \cosh\left(\frac{\pi\kappa v^2}{4}\right)}\\[2mm]
&=&1+6\pi^2 Q^4\left(35-3\nu^4\right)\kappa^2+\mathcal{O}(\kappa^3)
\,,\\[2mm]
\nonumber
B&=&\frac{2}{3\pi T_c v_c \cosh\left(\frac{\pi\kappa v^2}{4}\right)}\\[2mm]
&=& 1-18\pi^2 Q^4\left(7+\nu^4\right)\kappa^2+\mathcal{O}(\kappa^3)\\[2mm]
\nonumber
D&=&\frac{16Q^2}{\pi T_c v_c^3 \cosh\left(\frac{\pi\kappa v^2}{4}\right)}\\[2mm]
&=&1-18\pi^2 Q^4\left(15+\nu^4\right)\kappa^2+\mathcal{O}(\kappa^3)\,.
\end{eqnarray}
Equation~\eqref{lcs} has the same structure as the law of corresponding states 
for fluids~\cite{MannPT}
\begin{equation}
8\tau=3\nu\left(p+\frac{2}{\nu^2}\right)-\frac{1}{\nu^3}\,.
\end{equation}
It is easy to check that this equation is correctly  
restored for $\kappa\rightarrow0$, since $A=B=D=1$ in this limit.

\subsection{$P-v$ diagram}
\label{Pvdiagram}

\begin{figure}[t]
\begin{center}
\includegraphics[width=8cm]{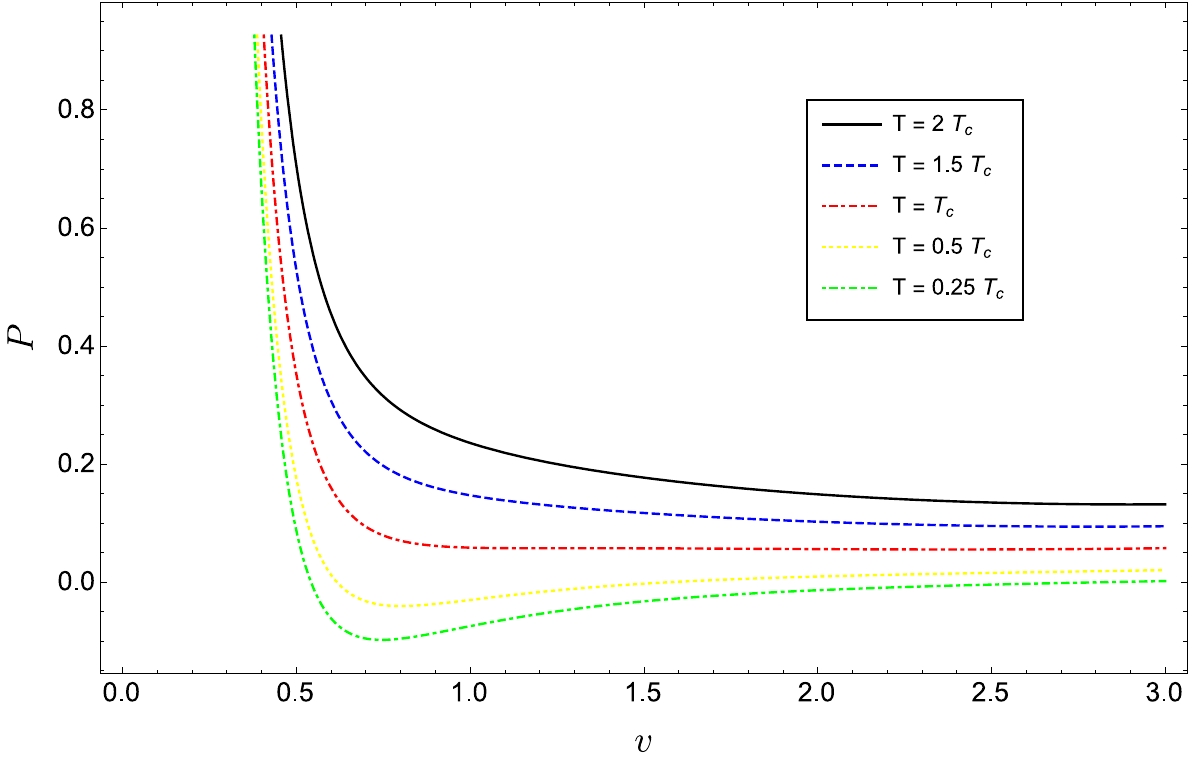}
\\[0.5cm]
\hspace{4mm}\includegraphics[width=8.0cm]{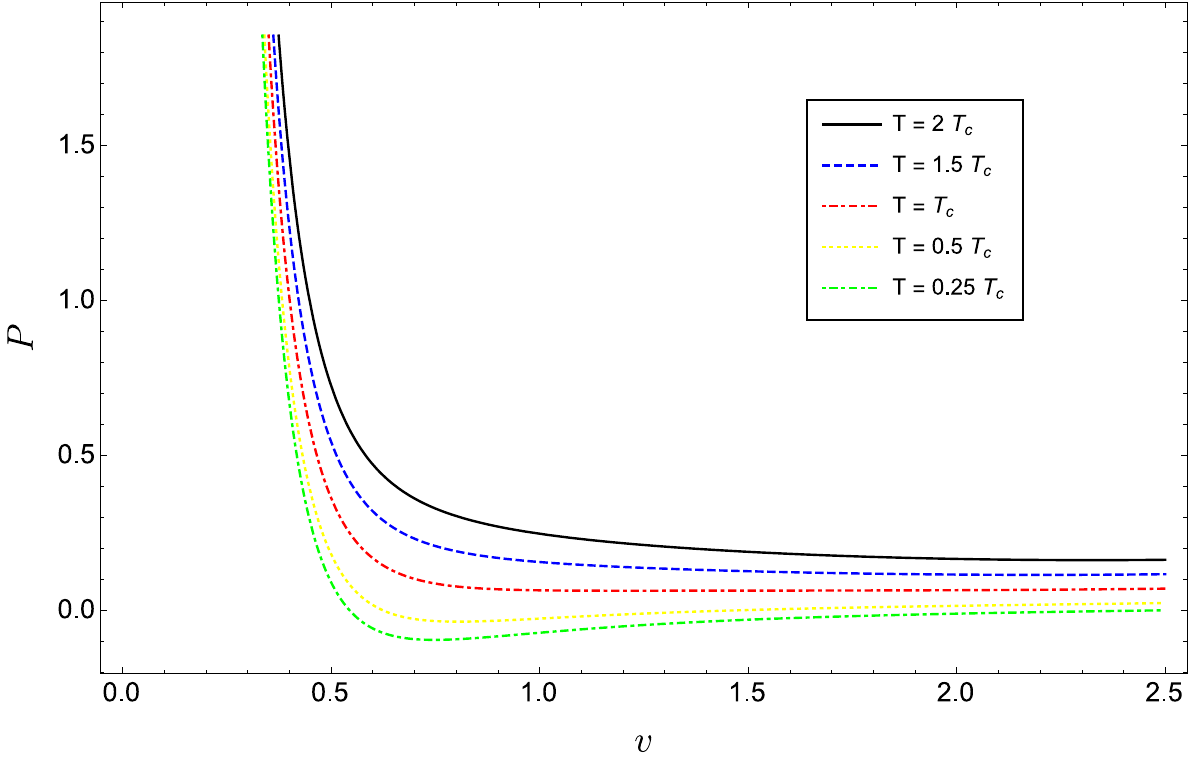}
\caption{{\it{$P-v$ diagrams for $\kappa=0.1$ (top panel) and $\kappa=0.15$ 
(bottom panel). In each panel, the temperature of the isotherms decreases from top to bottom. 
The red dot-dashed line indicates the critical isotherm at $T=T_c$ (online 
colors).}}}
\label{Fig6}
\end{center}
\end{figure}

We proceed by investigating the $P-v$ diagrams of AdS BHs as given in Eq.~\eqref{EoS2}. These diagrams are displayed in Fig.~\ref{Fig6}
for various $\kappa$, $T$ and fixed $Q$ as before. 
By comparison with Fig.~\ref{Fig1}, 
we see that the isotherms at $T<T_c$ (green and yellow curves) 
have van der Waals-like oscillations with a local minimum and maximum. As $T$ 
increases to $T_c$ (red curve), the oscillating branch squeezes and the two 
stationary points converge into the inflection point $(P_c,v_c, T_c)$ (see 
Eqs.~\eqref{vc}-\eqref{Pc}). 
This behaviour is reminiscent of the van der Waals fluid transition. Though not 
changing the qualitative behavior of the isotherms, Kaniadakis entropy 
non-trivially affects the 
critical pressure and temperature
at which such transition occurs.
For $T>T_c$ (blue and black curves), there are no more stationary points and $P$ 
decreases monotonically along each isotherm. 

\subsection{Gibbs  free energy}
\label{GE}

Let us now explore the global stability of charged AdS BHs in Kaniadakis 
thermodynamics. 
For this purpose, we compute
the Gibbs free energy as~\cite{MannPT,Deng:2018wrd}
\begin{eqnarray}
\label{Gibbs}
G(T,P)&=&M-TS\,=\,\frac{3\left(Q^2+r_+^2\right)+8\pi P r_+^4}{6r_+}\\[2mm]
\nonumber
&&+\,\frac{\left[Q^2-r_+^2\left(1+8\pi P r_+^2\right)\right]}{4\pi \kappa 
r_+^3}\tanh\left(\pi\kappa r_+^2\right)\,,
\end{eqnarray}
where $r_+$ is to be regarded as a function of $P$ and $T$ through  the equation 
of state~\eqref{EoS}.
Once more, one can check that the $\kappa\rightarrow0$ limit gives back the 
classical Gibbs free energy for charged AdS BHs 
\be
G_{\kappa\rightarrow0}(T,P)=\frac{3Q^2}{4r_+}+\frac{r_+}{4}-\frac{2}{3}\hspace{
0.2mm}\pi P r_+^3\,.
\ee

\begin{figure}[t]
\begin{center}
\includegraphics[width=8cm]{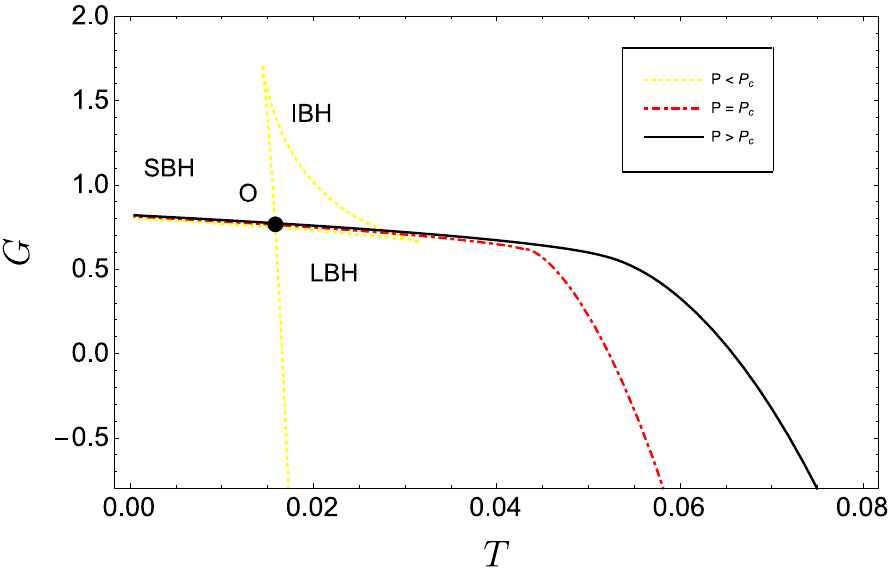}
\caption{{\it{The Gibbs  free energy $G$ versus the temperature $T$, for 
$\kappa=0.1$ (the same qualitative behavior is obtained for other values of 
$\kappa$). The red dot-dashed line indicates the critical isobar at $P=P_c$. SBH denotes  Region I - Small Black Hole, IBH denotes  Region 
II - Intermediate Black Hole and LBH denotes Region III - Large Black Hole, see text (online colors).
}}}
\label{Fig7}
\end{center}
\end{figure}

\begin{figure}[!]
\includegraphics[width=8cm]{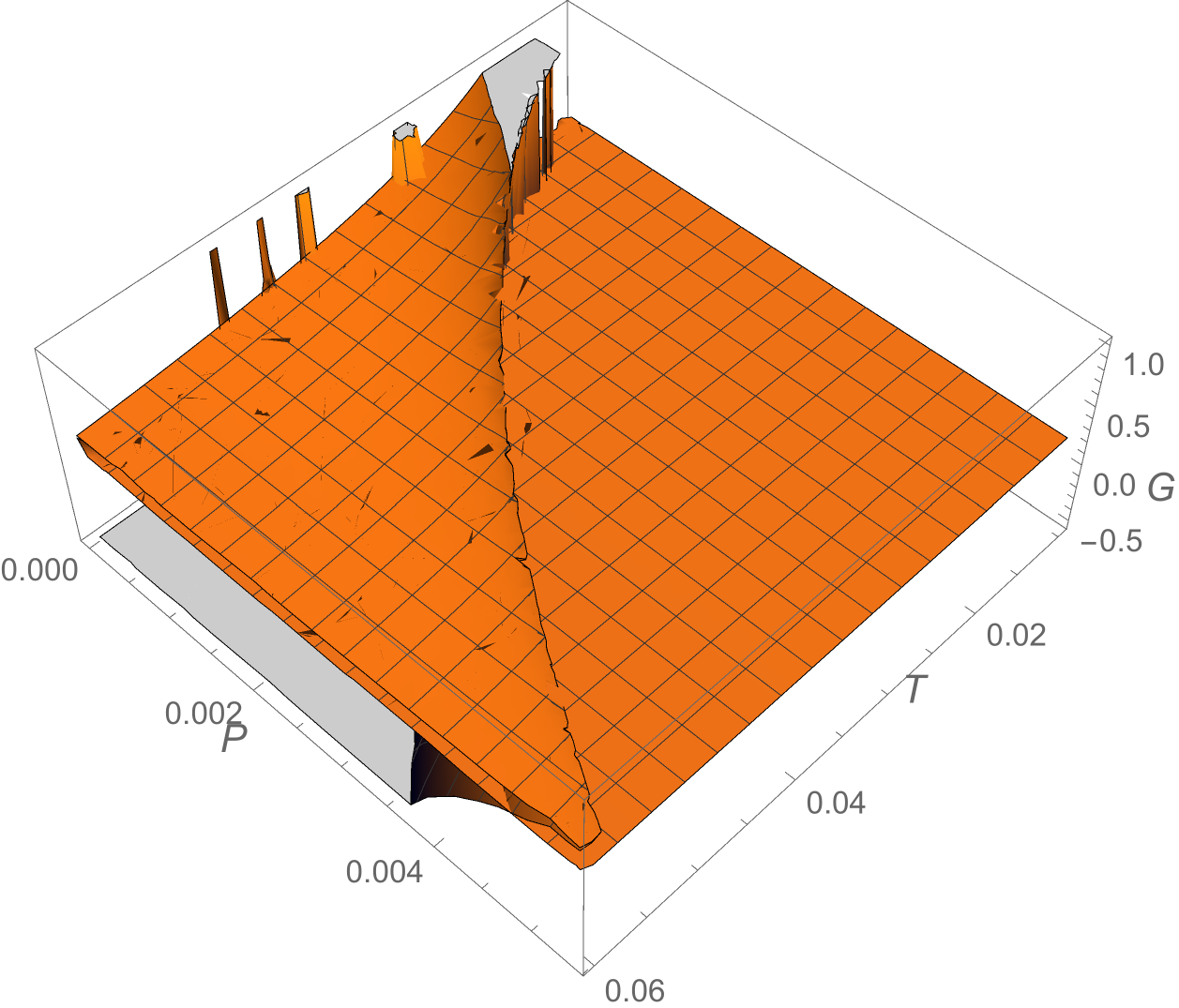}
\caption{{\it{3D plot of the Gibbs free energy $G$ versus pressure the $P$ and 
temperature $T$, for $\kappa=0.1$ (the same qualitative behavior can be obtained 
for other values of $\kappa$). The coexistence line between the SBH and LBH 
phases in the $P-T$ plane is visible.}}}
\label{Fig8}
\end{figure}

The behavior of Eq.~\eqref{Gibbs} as a function of $T$ is shown in 
Fig.~\ref{Fig7}, to be compared with Fig.~\ref{Fig2}. Consistently with the 
previous discussion,  it is observed that, below the critical pressure (dotted 
yellow line), $G$ exhibits has the swallow tail behavior typical of first order 
phase transitions. Specifically, in the first branch BHs 
are in the SBH domain. As $T$ increases to the critical point $O$, 
SBH and LBH phases coexist, since they 
have the same Gibbs free energy. 
As noted in~\cite{MannPT}, the coexistence line in the $P-T$ plane
can be derived by using Maxwell's equal area law or finding a curve for which 
$G$ and $T$ coincide for SBH and LBH. This line is plainly visible from the 3D 
plot in Fig.~\ref{Fig8}. 
Above the critical temperature, LBH 
becomes the preferred thermodynamic state because of its lower Gibbs free 
energy. Therefore, 
there is a first-order small-large
phase transition at the point $O$. Clearly, owing to the definition~\eqref{KEn} 
of entropy, different horizon areas for the SBH and LBH during this transition 
correspond to a discontinuity in the entropy (and also in the thermodynamic 
volume, see Eq.~\eqref{Vol}) and, thus, to the release of latent heat.

\subsection{Behavior near the critical point}
For quantitative discussion of the behavior of BHs approaching the critical 
point, we now calculate the critical parameters as defined at the end of 
Sec.~\ref{PVvdW}. First, we introduce the free energy
\begin{eqnarray}
F(T,V)&=&G-PV\,.
\end{eqnarray} 

By using Eqs.~\eqref{Vol} and~\eqref{Gibbs}, we get
\be
F=\frac{1}{2}\left[\frac{Q^2}{r_+}+r_+-\frac{2T\sinh\left(\pi \kappa 
r_+^2\right)}{\kappa}\right]\,.
\ee
Accordingly, the entropy is
\be
S(T,V)=-\left(\frac{\partial F}{\partial T}\right)_V=\frac{\sinh\left(\pi\kappa 
r_+^2\right)}{\kappa}
\,,
\ee
consistently with Eq.~\eqref{KEn}. In turn, from the definition below 
Eq.~\eqref{TildeQ}, we find that $C_V=0$, which yields $\alpha=0$.

In order to compute $\beta$, we approximate Eq.~\eqref{lcs} around a critical 
point. We use the re-scaled coordinates~\eqref{TildeQ}, here rewritten for 
convenience as
\be
t=\frac{T}{T_c}-1\,, \qquad \omega=\frac{V}{V_c}-1\,.
\ee
In the approximation of small $\kappa$, we obtain 
\begin{eqnarray}
\nonumber
p&=&1+\left(\frac{8}{3}-512\pi^2 Q^4\kappa^2\right)t + 
\left(-\frac{8}{3}+704\pi^2 Q^4\kappa^2\right)t\omega\\[2mm]
&&
+\left(-\frac{4}{3}+1120\pi^2 
Q^4\kappa^2\right)\omega^3+\mathcal{O}\left(t\omega^2,\omega^4\right)\,,
\label{pdiff}
\end{eqnarray}
where the various terms have been grouped together in this specific way for a direct 
comparison with~\cite{MannPT}. 
It is worth noting that the 
re-scaled pressure as appears in Eq.~\eqref{pdiff} contains terms of order 
higher than the qudratic in $\kappa$, due to the implicit $\kappa$-dependence of 
$t,\omega$. However, for our purpose of computing Kaniadakis corrections to 
critical exponents, it is useful to present Eq.~\eqref{pdiff} et seq. in their 
current form and restore the leading order at the end. For more details on the 
validity of the series expansion respect to $t,\omega$, see~\cite{MannPT}.

Now, differentiation of Eq.~\eqref{pdiff} respect to $\omega$ at a fixed $t<0$ 
gives
\begin{eqnarray}
dp&=&\\[2mm]
\nonumber
&&\hspace{-11mm}P_c\left[ \left(-\frac{8}{3}+704\pi^2 
Q^4\kappa^2\right)t+\left(-4+3360\pi^2 
Q^4\kappa^2\right)\omega^2\right]d\omega\,.
\end{eqnarray}
By employing Maxwell's equal area law
\be
\label{Max}
\oint v dP =0\,,
\ee
along with the knowledge that there is no pressure variation during the phase 
transition, we obtain 
\begin{widetext}
    \begin{eqnarray}
    \nonumber
&&1+\left(\frac{8}{3}-512\pi^2 Q^4\kappa^2\right)t + \left(-\frac{8}{3}+704\pi^2 
Q^4\kappa^2\right)t\omega_l
+\left(-\frac{4}{3}+1120\pi^2 Q^4\kappa^2\right)\omega_l^3\\[2mm]
&&= 1+\left(\frac{8}{3}-512\pi^2 Q^4\kappa^2\right)t + 
\left(-\frac{8}{3}+704\pi^2 Q^4\kappa^2\right)t\omega_s
+\left(-\frac{4}{3}+1120\pi^2 Q^4\kappa^2\right)\omega_s^3   \end{eqnarray}
and 
\be
0=\int_{\omega_l}^{\omega_s}\omega\left[ \left(-\frac{8}{3}+704\pi^2 
Q^4\kappa^2\right)t+\left(-4+3360\pi^2 
Q^4\kappa^2\right)\omega^2\right]d\omega\,,
\ee
\end{widetext}
where $\omega_{s,l}$ denote the ``specific volume'' of the small and large 
phases of BHs, respectively. One can verify that the only non-vanishing 
solution that reduces to the standard one for $\kappa\rightarrow0$ is~\cite{MannPT}
\be
\omega_s=-\omega_l=\sqrt{-2t}\left(1+288\pi^2 Q^4\kappa^2\right)\,.
\ee
Thus, from the definition of the critical exponent $\beta$ below 
Eq.~\eqref{TildeQ}, it follows that
\be
\label{crexpeta}
\eta=V_c\left(\omega_l-\omega_s\right)=2V_c\omega_l\propto 
\sqrt{-t}\,\,\, \Longrightarrow \,\,\,\beta=\frac{1}{2}\,.
\ee

As concerns the exponent $\gamma$, we need to differentiate Eq.~\eqref{pdiff} as
\begin{equation}
\left(\frac{dV}{dP}\right)_T=\frac{V_c}{P_c}\left(\frac{d\omega}{dp}
\right)_T=-\frac{3}{8}\frac{V_c}{P_c}\frac{1}{t}\left(1+264\pi^2 
Q^4\kappa^2\right).
\end{equation}
Hence, the isotherm compressibility $\kappa_T$ of BHs takes the form
\be
\kappa_T=-\frac{1}{V}\hspace{0.2mm}\left(\frac{\partial V}{\partial 
P}\right)_T\propto\frac{1}{t}\,,
\ee
which implies $\gamma=1$.

Lastly, the shape of the critical isotherm $t = 0$ and the related 
$\delta$-exponent are given by
\be
p-1=\left(-\frac{4}{3}+1120\pi^2 Q^4\kappa^2\right)\omega^3\,\,\, 
\Longrightarrow\,\,\, \delta=3\,.
\ee

In spite of the non-trivial modifications induced by the $\kappa$-deformed 
entropy to the critical pressure, volume and temperature, the basic critical exponents remain unaffected. This allows to conclude that the qualitative similarity between Kaniadakis BHs and van der Waals fluids near the critical point holds at quantitative level too. 

\subsection{Sparsity of black hole radiation}
\label{spars}

Although BHs nearly behave like a black body and spontaneously emit particles at 
a temperature proportional to their surface gravity, the flow of Hawking 
radiation exhibits some peculiar features. For instance, it is known to be more 
sparse than black body radiation.
Quantitatively speaking, such a difference can be estimated through the computation of the so-called sparsity, which is a measure 
of the average time-gap between the emission of successive quanta defined by
\be
\label{teta}
\tilde \eta=\frac{C}{g}\left(\frac{\lambda_t^2}{A_{eff}}\right),
\ee
(we have used the symbol $\tilde\eta$ instead of the traditional $\eta$ to avoid 
confusion with the critical exponent~\eqref{crexpeta}). 
Here, the constant $C$ is 
dimensionless, 
while $g$, $\lambda_t=2\pi /T$ and 
$A_{eff}=27A_{bh}/4$ denote the
spin degeneracy factor of the emitted quanta, the thermal wavelength and the 
effective horizon area of the BH, respectively. For Schwarzschild BHs and 
emission of massless bosons, one has $\lambda_t=2\pi/T_H=8\pi^2r_h$, which 
entails 
\be
\tilde \eta_H=\frac{64\pi^3}{27}\approx 73.49\gg1\,.
\ee
For comparison, we remind that 
$\tilde \eta\ll1$ in the case of a black body. 

Effects of deformed entropies on sparsity of Schwarzschild BHs have been 
recently considered in literature (see~\cite{Cimi,Alonso,Alonsobis,Rainb} and 
references therein). For instance, in~\cite{Cimi} it has been shown that 
generalized models of Heisenberg relation combined with non-extensive (R\'enyi, 
Tsallis-Cirto, Kaniadakis, Sharma
Mittal and Barrow) entropies lead to substantial modifications of the
sparsity, which turns out to be mass dependent. A similar statement has been 
claimed in~\cite{Rainb} in rainbow gravity. The question arises as to how such 
results appear for charged AdS BHs in Kaniadakis framework.

\begin{figure}[t]
\begin{center}
\includegraphics[width=8cm]{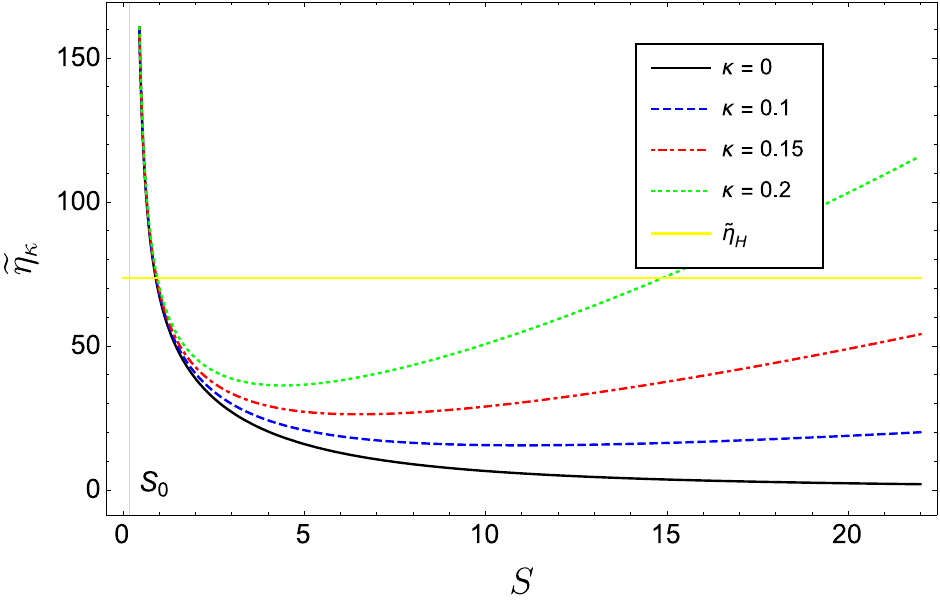}
\caption{{\it{The sparsity   $\tilde\eta_\kappa$ versus the entropy $S$, for 
various $\kappa$ and $l=2$. The vertical lines represent the physical limitation 
entropy $S_0$ for each curve. For comparison, we have also depicted the sparsity of 
Schwarzschild BHs with emission of massless bosons (yellow curve) (online 
colors).}}}
\label{Fig9}
\end{center}
\end{figure}

To compute the $\kappa$-deformed sparsity, we resort to the 
definition~\eqref{teta} equipped with Eq.~\eqref{T}. Straightforward 
calculations yield 
\be
\label{kspars}
\tilde\eta_\kappa=\tilde\eta_H\frac{\pi^2 l^4\kappa^2\left(1+\kappa^2 
S^2\right)\arcsinh^2\left(\kappa S\right)}
{\left[-\left(\pi l Q \kappa\right)^2+\pi l^2\kappa \arcsinh\left(\kappa 
S\right)+3\arcsinh^2\left(\kappa S\right)\right]^2}.
\ee
We notice that in the Schwarzschild limit (i.e. $Q=0$ and sufficiently large AdS 
radius $l$),
this expression reduces to
\be
\tilde\eta_\kappa (Q=0,l\rightarrow\infty)=\tilde\eta_H 
\left(1+\kappa^2S^2\right),
\ee
which coincides with the result of~\cite{Cimi} for Schwarzschild BHs. By further 
imposing $\kappa\rightarrow0$, we have $\tilde\eta_\kappa=\tilde\eta_H$, as 
expected. 

The behavior of the $\kappa$-modified sparsity~\eqref{kspars} for various 
$\kappa$ and fixed $l=2$, $Q=0.25$ is plotted in Fig.~\ref{Fig9}, which shows an 
apparent divergence for a certain ($\kappa$-dependent) value of $S$. This 
singularity, however, lies at the physical limitation point $S_0$ (see the 
discussion below Eq.~\eqref{T}), as it is easy to understand from the 
definition~\eqref{teta}. Thus, it is unphysical and we only have to consider the 
region $S>S_0$ delimited by the vertical line. We can see that the 
$\kappa$-deformed sparsity lies always above the $\kappa=0$ (black) curve, in 
such a way that increasing the value of $\kappa$ directly results
in sparser Kaniadakis BH radiation. This is in line with the result 
of~\cite{Cimi}. On the other hand, $\tilde\eta_\kappa$ is greater than the 
sparsity $\tilde\eta_H$ of Hawking radiation of Schwarzschild BHs (yellow 
curve) for sufficiently small and large entropies, where it significantly departs from 
the black body-like behavior, while it falls below in the intermediate range.

\section{Geometrothermodynamics of charged AdS black holes}
\label{Geom}

Since it is possible to define a temperature for BHs, it is natural to 
think of an associated
substructure. Recently, special care has been devoted to analyze the 
microscopic constituents and underlying interactions of 
BHs~\cite{Cai:1998ep,Wei:2015iwa,Wei:2019uqg,Guo:2019oad,Xu:2020gud,
Ghosh:2020kba,Xu:2020ftx,Prom,Dehghani:2023yph,Santos:2023eqp}, which can be 
described in the same fashion as the molecules of a non-ideal fluid. 

To investigate phenomenologically the nature of interactions among BH 
microstructures, the common technique consists in studying the thermodynamic 
geometry of the whole macroscopic system. In this perspective, the analysis of 
Weinhold~\cite{Wein1} and Ruppeiner~\cite{Rupp1,Rupp2}
geometries has proved to give qualitative insights on the internal dynamics of 
ordinary thermodynamics systems via exploring the sign of the corresponding 
metric curvature. 
Specifically, negative (positive) scalar curvatures emerge for prevailing attractive 
(repulsive)
microinteractions, while flatness characterizes non-interacting systems, such as 
the ideal gas, or systems where interactions are perfectly balanced.  

In the effort to probe the character of 
BH microinteractions, Weinhold and Ruppeneir formalisms have been adapted to BH 
thermodynamics. This kind of study has been first developed for Banados, 
Teitelboim and Zanelli (BTZ) BHs~\cite{Cai:1998ep} and later
extended to Reissner-Nordstr\text{\"o}m, Kerr and 
Reissner-Nordstr\text{\"o}m-AdS BHs~\cite{Shen:2005nu}. In the plethora of 
results obtained so far, there is general consensus that the scalar curvature of 
BH systems with charged molecules should be positive, revealing a repulsive 
behavior of 
microinteractions~\cite{Wei:2015iwa,Guo:2019oad,Wei:2019uqg,Ghosh:2020kba}. 

In order to figure out to what extent Kaniadakis' prescription~\eqref{KEn} affects the above conclusion, let us compute Weinhold and Ruppeneir scalar curvature in Kaniadakis 
entropy-based thermodynamics. Toward this end, we remind that
Weinhold metric is defined as the second derivative of internal energy of the 
system
with respect to given thermodynamic variables~\cite{Wein1}.
For the case of BHs, by identifying the internal energy with the mass, we 
obtain 
\begin{eqnarray}
\label{gw}
g_{ij}^w=-\partial_i\partial_j M(S,P,Q)\,\,\,\,\Longrightarrow\,\, 
\,\,ds^2_w=g^w_{ij} dx^i dx^j\,,
\end{eqnarray}
where we have we have generically denoted the 
independent fluctuation coordinates by $x^i$.

Similarly, in Ruppeiner formalism one considers 
the entropy as basic thermodynamic potential, i.e.
\be
\label{R1}
g_{ij}^{Rup}=-\partial_i\partial_j S\,.
\ee
From Eqs.~\eqref{T},~\eqref{gw} and~\eqref{R1}, 
it follows that the line elements of Weinhold and Ruppeiner are connected each 
other via the conformal transformation~\cite{Mrugala}
\be
\label{R2}
ds^2_R=\frac{ds^2_w}{T}\,.
\ee 

We focus our next geometrothermodynamic analysis on Eq.~\eqref{R2}. Considering 
the entropy and pressure as the fluctuation coordinates, while keeping $Q$ 
fixed, we obtain the following expression for the Ruppeiner scalar curvature: 
%\begin{widetext}
 %   \be
%\label{Rup}
%R^{Rup}(S,P)=\frac{\kappa^2\left[-2\pi Q^2\kappa+\arcsinh\left(\kappa 
%S\right)\right]}{\left(1+\kappa^2 S^2\right)^\frac{1}{2}\arcsinh\left(\kappa 
%S\right)\left\{\pi\left(Q\kappa\right)^2-\arcsinh\left(\kappa 
%S\right)\left[\kappa+8P\arcsinh\left(\kappa S\right)\right]\right\}}\,,
%\ee
%\end{widetext}
 \begin{eqnarray}
R^{Rup}(S,P)&\hspace{-1.5mm}=\hspace{-1.5mm}& 
\left(1+\kappa^2 S^2\right)^{-\frac{1}{2}} \left[
\arcsinh\left(\kappa 
S\right)\right]^{-1} \nonumber\\[2mm]
&&\times
\kappa^2\left[-2\pi Q^2\kappa+\arcsinh\left(\kappa 
S\right)\right]\nonumber\\[2mm]
&&\, \times
\left\{\pi\left(Q\hspace{0.2mm}\kappa\right)^2-\arcsinh\left(\kappa 
S\right)\left[\kappa\!+\!8P\arcsinh\left(\kappa S\right)\right]\right\}^{-1}\,,
  \label{Rup}
  \end{eqnarray}
which reduces to the standard curvature for charged AdS BHs
in the $\kappa\rightarrow0$ limit~\cite{Guo:2019oad}
\be
\label{Rlim}
R^{Rup}_{\kappa\rightarrow0}(S,P)=\frac{2\pi Q^2-S}{S\left[-\pi 
Q^2+S\left(1+8PS\right)\right]}\,.
\ee

 \begin{figure}[t]
\begin{center}
\includegraphics[width=8cm]{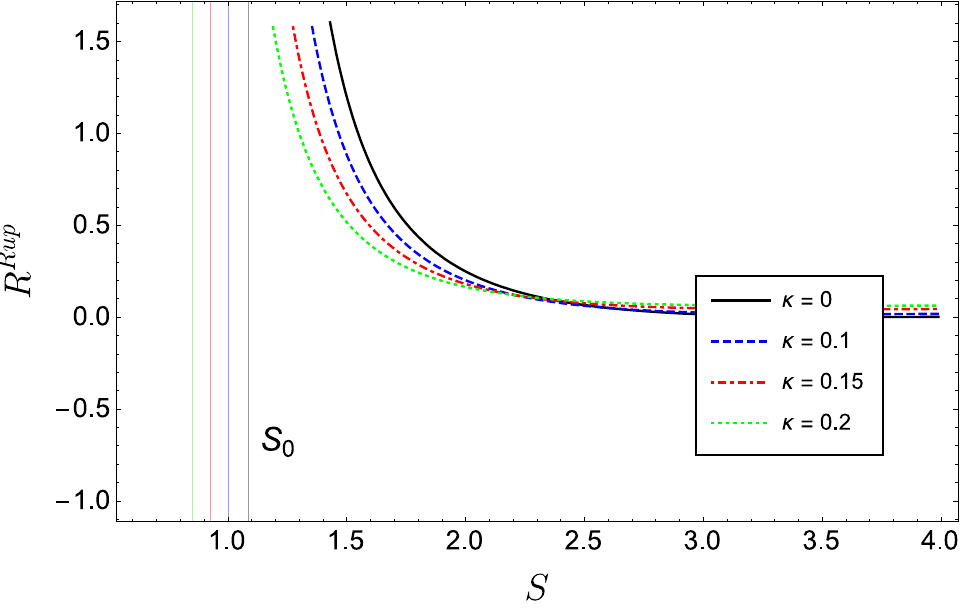}
\caption{{\it{The Ruppeiner scalar
curvature  $R^{Rup}$ versus the entropy $S$, for various values of $\kappa$. 
The parameter $l$ is fixed to $l= \sqrt{2} l_{c}$ defined through the critical 
condition~\eqref{Pc}. The vertical lines represent the physical limitation 
entropy $S_0$ for each curve 
(online colors).}}}
\label{Fig10}
\end{center}
\end{figure}

The curvature~\eqref{Rup} versus $S$ is displayed in Fig.~\ref{Fig10} for 
various $\kappa$ and fixed $Q=0.6$, $P=0.5 P_c$. As discussed for the sparsity above, 
the physical region is delimited by  $S>S_0$ (vertical lines). Although Eq.~\eqref{Rup} is 
non-trivially modified comparing to the classical curvature~\eqref{Rlim},  
Kaniadakis entropy does not affect the overall sign of $R^{Rup}$, which is still 
positive and indicates
prevailing repulsive interactions among BH microstructures. 
As $S$ increases, $R^{Rup}$
gradually decreases,
which means that the repulsion progressively fades, possibly due to 
thermal fluctuations and/or molecular collisions.
Kaniadakis corrections here manifest through a variation of the rate of 
decrease, with higher $\kappa$ corresponding to faster decrease for sufficiently 
small $S$, and vice-versa.   This tendency is reversed for $S$ large enough. The 
former behavior resembles the physics of composite systems with non-extensive 
(and, in particular, superadditive) entropy. Indeed, for such systems the 
single 
constituents tend to merge more strongly than the classical extensive 
case~\cite{Vedral}, thus 
balancing swiftly the effects of internal repulsive forces. 
Asymptotically (i.e. for large BH horizon radii), the internal 
microstructures end up being so far apart that $R^{Rup}\rightarrow0$, 
which reveals that BHs 
behave as effectively non-interacting.

\begin{figure}[t]
\begin{center}
\includegraphics[width=8cm]{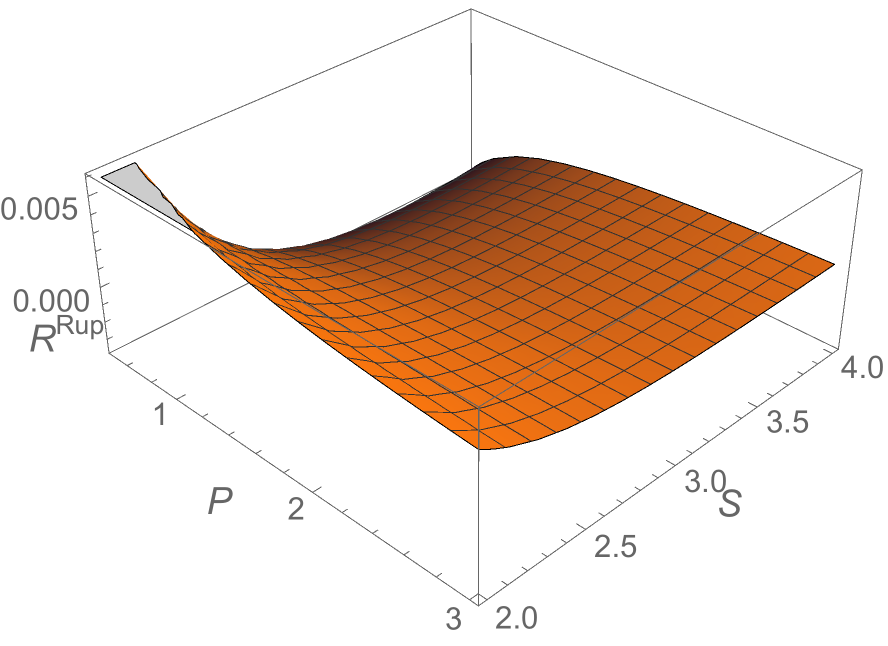}
\caption{{\it{3D plot of the Ruppeiner scalar
curvature  $R^{Rup}$ versus entropy $S$ and pressure $P$, for $\kappa=0.2$ and 
$Q=0.6$.}}}
\label{Fig11}
\end{center}
\end{figure}

Finally, Fig. \ref{Fig11} shows the 3D plot of $R^{Rup}$ versus $S$ and $P$ 
for $\kappa=0.2$ and fixed $Q$ as before.  We can see that the scalar curvature 
remains positive 
even for varying $P$, which 
supports previous arguments on the repulsive nature of BH micro-interactions. 

\section{Conclusions and Discussion}
\label{Disc}

Geometrothermodynamics and phase transitions of charged AdS BHs have been addressed within the framework of Kaniadakis theory, which arises from a self-consistent relativistic generalization of the classical statistical 
mechanics. The latter is coherently recovered by setting the deformation 
parameter $\kappa$ to zero. We would like to stress that the highlight of the 
present analysis is to deepen our knowledge of BH thermodynamics in a fully 
relativistic statistical scenario. As far as we know, this is the first work where  
this scenario is addressed. 

Following the standard literature, the study has been conducted by identifying 
the cosmological constant and its conjugate quantity with the thermodynamic 
pressure and volume, respectively. In the ensuing extended phase space, we have 
examined the impact of Kaniadakis entropy on the formal duality 
black-hole/fluid, showing that Kaniadakis BHs still exhibit a van der Waals-like 
first order phase transition.

Although Kaniadakis corrections do not affect the qualitative behavior of $P-v$  
diagrams and the basic critical exponents, the critical volume, pressure and 
temperature are non-trivially modified, even at the leading order in the 
deformation parameter $\kappa$ (see Eqs.~\eqref{vc}-\eqref{Pc}). Should we have 
access to the phenomenology of AdS BHs and measure such quantities, we could 
elaborate more on the role of Kaniadakis entropy in BH physics
and possibly constrain $\kappa$-corrections. We have finally probed the nature of 
interactions among BH micro-structures. Using the picture of 
fluid-like interacting molecules, we have applied Ruppeiner 
geometrothermodynamic formalism and computed the scalar curvature of the 
associated metric. The investigation of the sign of the Ruppeiner scalar 
curvature $R^{Rup}$ reveals that 
these micro-interactions are prevailing repulsive and tend to vanish for 
sufficiently large BH horizon radii, with the $\kappa$-parameter ruling the rate 
of decrease. In passing, we mention that a possible explanation for this 
behavior can be provided based on the physics of the two fluid model, where the 
dominant character of interactions is determined by the relative number 
densities of the molecules of the two fluids~\cite{Guo:2019oad}.

As future prospects, it would
be interesting to enrich the above analysis by considering the presence  of 
global monopoles, which are known to have non-trivial effects on BH 
physics~\cite{Jing:1993np,Yu:1994fy,Li:2002ku,Jiang:2005xb,Carames:2017ngt}.
Furthermore, one can study Kaniadakis entropy-based thermodynamics of other BHs, 
such as rotating or exotic BTZ BHs,  and additionally examine its effect on the 
primordial black holes and stochastic gravitational waves 
\cite{Papanikolaou:2022did,Basilakos:2023xof}. On the other hand, inspired 
by~\cite{Cimi}, it is suggestive to understand how BH critical phenomena appear 
in the context of modified uncertainty principles~\cite{Mignemi:2009ji,LucEUP}  
combined with nonextensive entropies, and possibly connect the two frameworks.  
The study of these aspects 
is under active consideration and will be presented elsewhere.

\acknowledgements
GGL would like to thank Jaume Gin\'e for useful comments on the original manuscript. He is also grateful to the Spanish ``Ministerio de Universidades'' for the awarded Maria Zambrano fellowship and funding received from the European Union - NextGenerationEU.
The authors acknowledge the contribution of the LISA CosWG and of COST Actions  CA18108 ``Quantum Gravity 
Phenomenology in the multi-messenger approach''  and 
CA21136 ``Addressing observational tensions in cosmology with systematics and fundamental physics (CosmoVerse)''.

\end{document}